\documentclass[prb,aps,amsmath,amssymb,reprint,superscriptaddress,showpacs]{revtex4-1}
\usepackage{graphicx}
\usepackage[colorlinks=true,linkcolor=blue,filecolor=blue,urlcolor=blue]{hyperref}
\usepackage{epstopdf}
\pdfoutput=1 
\usepackage{color}

\newcommand{\calV}{\mathcal{V}}

\newcommand{\ua}{\uparrow}
\newcommand{\da}{\downarrow}

\begin{document}

\title{Selection of factorizable ground state in a frustrated spin tube: Order by disorder and hidden ferromagnetism}

\author{X. Plat}
\affiliation{Laboratoire de Physique Th\'eorique, Universit\'e de Toulouse and CNRS, UPS (IRSAMC), F-31062, Toulouse, France}
\author{Y. Fuji}
\affiliation{Institute for Solid State Physics, University of Tokyo, Kashiwa 277-8581, Japan}
\author{S. Capponi}
\affiliation{Laboratoire de Physique Th\'eorique, Universit\'e de Toulouse and CNRS, UPS (IRSAMC), F-31062, Toulouse, France}
\author{P. Pujol}
\affiliation{Laboratoire de Physique Th\'eorique, Universit\'e de Toulouse and CNRS, UPS (IRSAMC), F-31062, Toulouse, France}

\date{\today}


\begin{abstract}
The interplay between frustration and quantum fluctuation in magnetic systems is known to be the origin of many exotic states in condensed matter physics. 
In this paper, we consider a frustrated four-leg spin tube under a magnetic field.
This system is a prototype to study the emergence of a nonmagnetic ground state factorizable into local states and the associated order parameter without quantum fluctuation, that appears in a wide variety of frustrated systems. 
The one-dimensional nature of the system allows us to apply various techniques: a path-integral formulation based on the notion of order by disorder, strong-coupling analysis where magnetic excitations are gapped, and density-matrix renormalization group. 
All methods point toward an interesting property of the ground state in the magnetization plateaus, namely, a quantized value of relative magnetizations between different sublattices (spin imbalance) and an almost perfect factorization of the ground state. 
\end{abstract}

\maketitle

\section{Introduction}\label{sec:introduction}

Frustrated magnetism is a subject that has attracted much attention in the last decades. 
From the quantum mechanical perspective, frustration is the key element in the search
of exotics ground states, like spin liquids.~\cite{[See for instance the review by ][and references therein.]Balents2010} Very often, low-energy effective models, such as quantum dimer models, are used to get a better understanding of the physics of such frustrated systems. In non-bipartite lattices, they provide maybe the most controllable examples of states that can be assimilated to spin liquids~\cite{Moessner2001,Balents2002,Misguich2002} for a finite range of parameters, while for bipartite lattices only the Rokhsar-Kivelson point displays a disordered (critical) state.~\cite{Rokhsar1988} These exotic spin liquid states have the interesting property of topological degeneracy, which cannot be identified with a local order parameter. It is related to a long-range entanglement of the ground state \cite{Kitaev2006,Levin2006}. 

From the classical statistical mechanics perspective, frustrated systems have attracted also a lot of interest because of the phenomenon of order by disorder (OBD).~\cite{Villain1980} It is by now well understood that OBD is the mechanism that gives rise to a ground-state selection among a continuously degenerate manifold in classical frustrated magnets such as the Heisenberg model on the Kagom\'{e}~\cite{Chalker1992,Huse1992,Reimers1993,Cabra2002a,Zhitomirsky2002} or the pyrochlore \cite{Moessner1998,Moessner1998a,Palmer2000,Chern2008} lattices. Such systems present ``soft modes'' in their spin-wave spectra, and the configurations with the most soft modes will be favored entropically at low but non-zero temperature, against configurations with the same energy but less soft modes.~\cite{Henley1989,Shender1982} The straightforward extension of the ideas of OBD to quantum mechanics is simply to argue that, among many configurations with the same classical energy, the one that has the lowest zero-point energy quantum correction is selected, and a wide number of models on different lattices have been studied in this way~\cite{Cabra2002a,Henley1989,Chubukov1992,Tchernyshyov2003,Henley2006,Bergman2007,Canals2008,Coletta2011,Zhitomirsky2012,Henry2012} or sometimes going beyond harmonic level if needed.~\cite{Hizi2006,Gregor2006,Hizi2007,Hizi2009}

In this paper, we are going to argue that the phenomenon of classical OBD may be revealed in another and more subtle way. The symptoms of OBD that we discuss here can in some sense be found in the existing literature although they have not been enough emphasized in our opinion. More interestingly, they go somehow in the opposite direction of long-range entanglement in topological gapped quantum spin liquids.~\cite{Jiang2012a,Jiang2012b,Depenbrock2012,Nishimoto2013} Indeed, high frustration may lead to ground-state wave functions that are, to a large extent, factorizable into local states (i.e. a product state). The work of Schulenburg {\it et al}.~\cite{Schulenburg2002} for the Kagom\'{e} lattice provides an exact result in which highly frustrated magnets in the presence of a strong magnetic field have a factorizable wave function consisting in a collection of localized magnons when the system is close to saturation. This was later shown to be also the case for several lattices like the sawtooth chain, the checkerboard and pyrochlore lattices.~\cite{Richter2004} Although there is no exact result, the Heisenberg model on the Kagom\'{e} lattice at its ${1 \over 3}$ plateau is expected to have a wave function with a large overlap to a factorizable toy wave function.~\cite{Cabra2005,Capponi2013} A consequence of this is the fact that the relative magnetization between different sites of the lattice, or spin imbalance, is locked to a fixed value; for example, in the case of the Kagom\'{e} lattice, the total magnetization of a resonating hexagon, is fixed to an integer value as compared to the magnetization of the fully polarized spins surrounding the hexagon. 

To provide a better and more concrete understanding of the statements above, we are going to study the system which is certainly the simplest prototype to reveal such an interesting phenomenology: the frustrated four-leg spin tube. A first argument in favor of this system is its one-dimensional (1D) nature allowing to use powerful nonperturbative analytical and numerical techniques. Obviously, the chosen system gives rise at the classical level to the phenomenon of OBD and, as we are going to show, produces in a quite explicit way all the phenomenology we have mentioned above: a mechanism to lead the factorization of the ground state and the quantization of the order parameter. A second argument in favor of this model is the fact that it possesses magnetization plateaus, a common consequence of frustration but not necessary the most interesting one for the point we want to make here. Indeed, it is interesting to locate in a magnetization plateau and then focus on the fate of nonmagnetic excitations, which are going to be the principal actors of the desired physics. Last but not least, the four-leg tube is somehow the parent system of the three-leg spin tube, which has been extensively studied (see for example Ref.~\onlinecite{Sakai2010} and references therein) and is also a frustrated system showing the presence of magnetization plateaus. However, it does not have classical OBD and therefore does not give rise to the phenomenology in which we are interested here. It will play the role of a reference example to compare our results.

\begin{figure}
\includegraphics[clip,width=0.45\textwidth,clip]{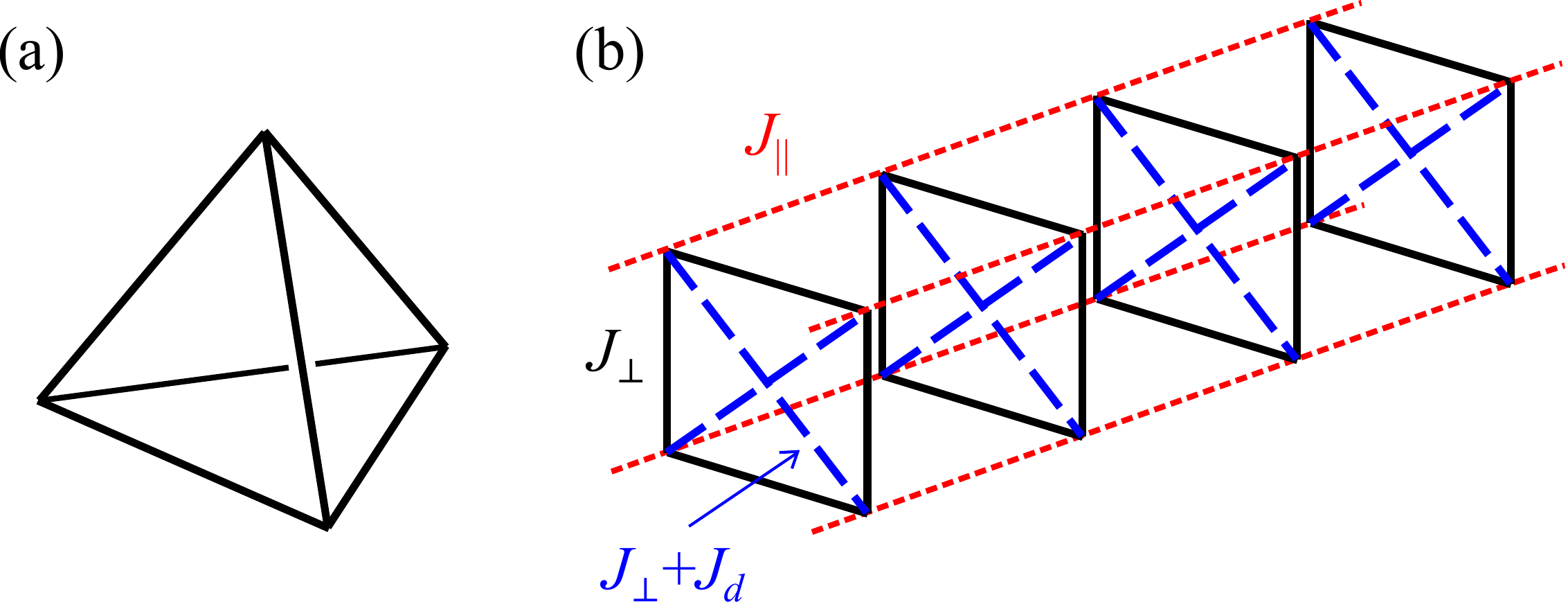}
\caption{(Color online) Schematic pictures of (a) a single-rung tetrahedron and (b) the four-leg spin tube composed of the coupled tetrahedra.}
\label{fig:Model}
\end{figure}

The Hamiltonian of the frustrated four-leg spin tube with diagonal couplings on the rungs and in a magnetic field is
\begin{eqnarray} \label{eq:4leg_tube_Ham}
H &=& J_{\parallel} \sum_{i=1}^L \left[ \sum_{j=1}^4 \vec{S}_{i,j} \cdot \vec{S}_{i+1,j} + J_\perp \sum_{j<j'} \vec{S}_{i,j} \cdot \vec{S}_{i,j'} \right. \nonumber \\ 
&& \left. + J_d \left( \vec{S}_{i,1} \cdot \vec{S}_{i,3} +\vec{S}_{i,2} \cdot \vec{S}_{i,4} \right) -h \sum_{j=1}^4 S^z_{i,j} \right], 
\end{eqnarray}
where $\vec{S}_{i,j}$ is the spin-1/2 operator on rung $i$ and on leg $j$, $L$ is the tube length, $J_{\parallel}$, 
$J_{d}$, and $J_{\perp}$ are positive antiferromagnetic couplings, and $h$ is a magnetic field along the $z$ axis. 
In Fig.~\ref{fig:Model}, we show a single-rung tetrahedron and the four-leg spin tube composed of the coupled tetrahedra. 
In this paper, we focus on the ground state properties of this model on several magnetization plateaus with fixed magnetization per site $m$: 
\begin{eqnarray}
m = \frac{1}{4L} \sum_{i=1}^L \sum_{j=1}^4 \langle S^z_{i,j} \rangle, 
\end{eqnarray}
where $\langle \cdots \rangle$ denotes the ground-state expectation value. 

At $J_d=0$, this model has a tetrahedral point-group symmetry $T_d$, or equivalently, a permutation symmetry $S_4$ of the four chains. 
Regarding this symmetry, it may share some common properties with the three-dimensional (3D) pyrochlore lattice. 
Since our model is in 1D and strongly frustrated, we have a particular interest in its nonmagnetic properties.  
Such nonmagnetic features naturally emerge in the pyrochlore lattice built on coupled tetrahedra, since the triplet excitations are fully gapped in the decoupled limit. 
This model has been originally studied in the theoretical literature \cite{Harris1991,Canals1998,Tsunetsugu2001a,Tsunetsugu2001b,Koga2001,Berg2003,Kotov2005} but recently proposed experimentally.~\cite{Okamoto2013,Kimura2014} 
Although our model is apparently far from the experimental realization, it is easily tractable and then will be a simplest starting point to explore those 3D candidates in the presence of a magnetic field. 
Another remarkable feature of this model is the exact macroscopic degeneracy of the disordered ground state at the quantum level even after introducing tiny couplings between tetrahedra. 
In fact, this model can be mapped onto an SU(2) ferromagnet. 
By perturbing this ``hidden'' ferromagnet with additional couplings, a factorizable nonmagnetic ordered state is selected, as expected from our discussion about the OBD mechanism. 
We finally mention that another model of the frustrated four-leg spin tube has been studied recently.~\cite{Arlego2011,Arlego2013}

The paper is organized as follows. 
In Sec.~\ref{sec:Path_integral}, we use a large-$S$ path-integral approach to discuss the OBD phenomenon with the computation of the zero-point energy. 
We propose the emergence of quantized spin-imbalance phases. 
Then we consider in Sec.~\ref{sec:Strong_coupling} the strong-coupling limit of the model in certain magnetization plateaus and analyze the effective Hamiltonian. 
In Sec.~\ref{sec:num}, we compare our predictions to density-matrix renormalization group (DMRG) simulations. 
Sec.~\ref{sec:conclusion} is devoted to the summary of our results and conclusion. 
In Appendix~\ref{sec:appendixA}, several details on the strong-coupling analysis are supplemented. 

%%%%%%%%%%%%%%%%%%%%%%%%%%%%%%%%%%%%%%%%%%%%%%%%%%%%%%%%%%%%%%%%%%%%%%%%%%%%%%%%%%%%%%%%%%
%%%%%%%%%%%%%%%%%%%%%%%%%%%%%%%%%%%%%%%%%%%%%%%%%%%%%%%%%%%%%%%%%%%%%%%%%%%%%%%%%%%%%%%%%%
%%%%%%%%%%%%%%%%%%%%%%%%%%%%%%%%%%%%%%%%%%%%%%%%%%%%%%%%%%%%%%%%%%%%%%%%%%%%%%%%%%%%%%%%%%

\section{Path-integral analysis}\label{sec:Path_integral}
In this section, we present a semi-classical analysis of the model (\ref{eq:4leg_tube_Ham}) and show the occurrence of 
a ground-state selection by an OBD mechanism.~\cite{Villain1980} Indeed, we will see that the classical ground state of 
this model is continuously degenerate with the presence of a free angle variable. An important question is then to know which value 
of this angle is selected by the quantum fluctuation, or alternatively by the thermal fluctuations. It turns out that, in 
our case, these two kinds of fluctuation seems to act in a different manner. We finally discuss how the question of the tunneling between the different favored states arise and its consequences.

%%%%%%%%%%%%%%%%%%%%%%%%%%%%%%%%%%%%%%%%%%%%%%%%%%%%%%%%%%%%%%%%%%%%%%%%%%%%%%%%%%%%%%%%%%
%%%%%%%%%%%%%%%%%%%%%%%%%%%%%%%%%%%%%%%%%%%%%%%%%%%%%%%%%%%%%%%%%%%%%%%%%%%%%%%%%%%%%%%%%%

\subsection{Method}\label{sec:PI_method}

We follow a method recently developed by Tanaka, Totsuka, and Hu.~\cite{Tanaka2009} 
They used a Haldane's path-integral approach based on the spin coherent state~\cite{Klauder1979}. 
Haldane's analysis leads to an action comprising two terms.~\cite{Haldane1983} One is the coherent-state expectation value of the Hamiltonian, or simply the
Hamiltonian for the classical configuration. The other term is the Berry phase one and
corresponds to the surface area (or the solid angle), $\int d\tau [1-\cos \theta(\tau)]\partial_{\tau}\varphi(\tau)$ 
in spherical coordinates, enclosed by each spin during its imaginary-time 
$\tau$ evolution. 

In order to build a low-energy effective theory from this starting point, one proceeds by first identifying the classical solution,
\begin{equation}
\vec{S}_{i,j}=S\left( \sin \theta^{(0)}_{i,j} \cos \varphi^{(0)}_{i,j}, 
\sin \theta^{(0)}_{i,j} \sin \varphi^{(0)}_{i,j}, \cos \theta^{(0)}_{i,j} \right), 
\end{equation}
and then adding the quantum fluctuation on top of it, 
\begin{equation}
\begin{split}
\theta^{(0)}_{i,j} &\rightarrow \theta_{i,j}= \theta^{(0)}_{i,j}+ \delta \theta_{i,j}, \\ 
\varphi^{(0)}_{i,j} &\rightarrow \varphi^{(0)}_{i,j}+\varphi_{i,j}. 
\end{split}
\end{equation}
We then expand the spin components up to second order in $\delta \theta$. 
The calculation of the SU(2) commutation relations $[S_{i,j}^z,S_{k,l}^{\pm}] = \pm S^\pm_{i,j} \delta_{ik} \delta_{jl}$ 
leads to the new set of variables $\Pi_{i,j}$, defined by
\begin{equation}
\Pi_{i,j}=-S\left[\sin \theta^{(0)}_{i,j}\delta\theta_{i,j}+\frac{1}{2}\cos \theta^{(0)}_{i,j}\delta\theta_{i,j}^2\right],
\label{eq:PI_definition_pi_momentum}
\end{equation}
which are the conjugate momenta to the angular variables, $[\varphi_{i,j},\Pi_{k,l}]=i\delta_{ik}\delta_{jl}$. It ensures to
have the correct commutators for the spin operators. Then we rewrite
these operators as functions of the conjugate fluctuation variables, 
\begin{equation}
\begin{split}
S^{\pm}_{i,j}& = e^{\pm i\left[\varphi^{(0)}_{i,j}+\varphi_{i,j}\right]}S\bigg[\sin \theta^{(0)}_{i,j} -
\frac{m}{S\sin \theta^{(0)}_{i,j}}\Pi_{i,j}, \\
&-\frac{1}{2}\frac{S^2}{S^2-m^2}\frac{1}{S\sin \theta^{(0)}_{i,j}}\Pi_{i,j}^2\bigg]\\ 
S^z_{i,j}& = S \cos \theta^{(0)}_{i,j}+\Pi_{i,j}. 
\end{split}
\label{eq:PI_spin_operators}
\end{equation}
Inspecting the expression of $S^z_{i,j}$, it is clear that $\Pi_{i,j}$ represents the fluctuation around the classical magnetization per site, $m_{i,j}=S\cos \theta^{(0)}_{i,j}$ .
The action is then rewritten in a function of these variables at the second order.

%%%%%%%%%%%%%%%%%%%%%%%%%%%%%%%%%%%%%%%%%%%%%%%%%%%%%%%%%%%%%%%%%%%%%%%%%%%%%%%%%%%%%%%%%%
%%%%%%%%%%%%%%%%%%%%%%%%%%%%%%%%%%%%%%%%%%%%%%%%%%%%%%%%%%%%%%%%%%%%%%%%%%%%%%%%%%%%%%%%%%

\subsection{Classical ground state}\label{sec:PI_classical_gs}

\begin{figure}
\includegraphics[width=0.46\textwidth,clip]{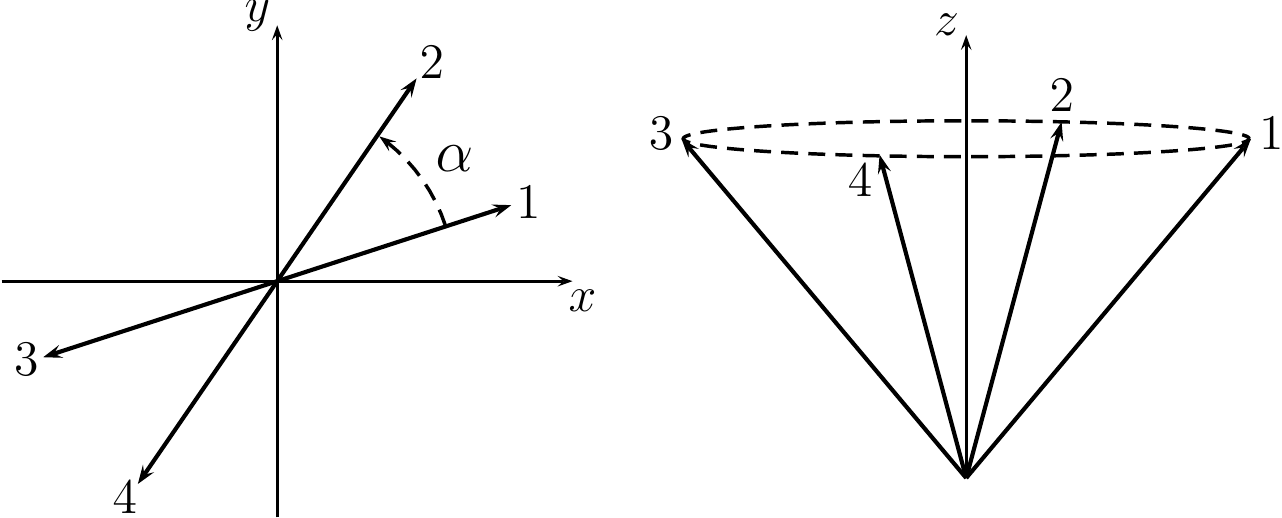}
\caption{Classical ground state of the model (\ref{eq:4leg_tube_Ham}). The four spins make a total spin zero in the $xy$ plane perpendicular to the field, where a 
free angle $\alpha$ is present (left panel), and are equally canted along the magnetic field in the $z$ direction (right panel).}
\label{fig:PI_classGS}
\end{figure}

From now on we focus on the regime $J_d\geq 0$. For $J_{\parallel}=0$, $J_d=0$, and $h=0$, the ground state on a rung is determined by the unique condition
$\vec{S}_{\boxtimes}=\vec{0}$ where $S^\mu_\boxtimes = \sum_{j=1}^4 S^\mu_j$, $\mu=x,y,z$. 
This leads to a continuous degeneracy of two angles in each rung. This is the same situation as the pyrochlore lattice as both systems share the same 
elementary cell.~\cite{Moessner1998}
If we add the magnetic field, there is the additional magnetization condition $S_{\boxtimes}^z=m$ and only one of the two angles remains free. The ground state is then given
by equally canting the four spins along the field and by making two pairs of antiparallel spins in the perpendicular $xy$ plane (Fig.~\ref{fig:PI_classGS}). 
The energy is independent of the angle $\alpha$ between the two spins 1 and 2 projected onto the $xy$ plane, thus in the decoupled rung limit we have one free angle per rung.

Coupling the rungs with a non-zero $J_{\parallel}$, only one free angle remains while we can accommodate a $k_{\parallel}=\pi$ 
state along the chain for the spin components in the $xy$ plane ($k_\parallel$ is the momentum along the chain). This angle is nothing but the continuous degeneracy 
that we discussed above. Therefore on top of the usual $\mathrm{U}(1)$ symmetry, we end 
up with an extra continuous degeneracy for the classical ground state. We want to emphasize that, because this angle is 
not associated with the symmetry of the Hamiltonian, we expect the quantum and thermal fluctuations to necessarily select, through an 
OBD mechanism, somes states with the corresponding angles minimizing the free energy of the system. 

We parametrize the classical solution with $\varphi^{(0)}_{i,1} = i\pi, \varphi^{(0)}_{i,2} = \alpha+ i\pi, \varphi^{(0)}_{i,3} = (i+1)\pi, \varphi^{(0)}_{i,4} = \alpha+ (i+1)\pi$, 
and $\cos \theta^{(0)}_{i,j} = h/[2S(2J_{\parallel}+2J_{\perp}+J_d)]$. It is important to note that by chosing such a parametrization we have broken 
a $\mathbb{Z}_3$ symmetry. Indeed, we are at the point $J_d=0$ with the $S_4$ symmetry group. 
Thus, instead of choosing sites 1 and 3 to be antiparallel as we did here, we could have chosen any 
of the three spins 2, 3, or 4 to be paired with the spin 1, that we consider as fixed. Once this choice is 
made, let us comment briefly about some differences depending on the value of $\alpha$, which has a periodicity of $2\pi$ and that we define as the angle between spins 1 and 2. 
From Fig.~\ref{fig:PI_classGS}, we see that the cases $\alpha=0,\pi$ and $\alpha=\pi/2,3\pi/2$ lead both to two distinct states, while for a generic value of $\alpha$ there are four non-equivalent 
states with values of the angle between spins 1 and 2 taking the values $\alpha,\pi-\alpha,\pi+\alpha,2\pi-\alpha$. 

To distinguish between those four states, we propose to use the following couple of operators
\begin{equation}
\begin{split}
\chi^{1234} &= \sum_{j=1}^4 \left( \vec{S}_j \times \vec{S}_{j+1} \right)^z, \\ 
Q^{1234} &= \left( \vec{S}_{1} \times \vec{S}_{2} \right) \cdot \left( \vec{S}_{3} \times \vec{S}_{4} \right).
\end{split}
\label{eq:PI_QChi_operators}
\end{equation}
The operator $\chi^{1234}$ is the usual measure of the $z$-component of the spin vector chirality, and the $Q^{1234}$ operator is discussed in the strong-coupling analysis of 
Sec.~\ref{sec:Strong_coupling} where these operators will be of great use. Computing their values in 
the four states for a generic $\alpha$, we get 
\begin{equation}
\begin{split}
\langle \chi^{1234} \rangle &\sim \left( \sin (\alpha), \sin (\alpha), -\sin(\alpha), -\sin(\alpha) \right), \\
\langle Q^{1234} \rangle &\sim \left( \mathrm{q}_{-}(\alpha), \mathrm{q}_{+}(\alpha), \mathrm{q}_{+}(\alpha), \mathrm{q}_{-}(\alpha) \right),
\end{split}
\end{equation}
where 
\begin{equation}
\mathrm{q}_{\pm}(\alpha) =  (2m^2-S^2) \sin^2(\alpha) + m^2 \left(1 \pm \sin(\alpha) \right)^2, 
\end{equation}
and those are always nonzero.

When $J_d>0$, those states remain ground states. The only difference is that we do no longer have the three 
possibilities when anti-aligning a spin with the spin 1, thus the above discussion also applies to this regime.

Finally, we discuss the symmetry relations between the four states when $\alpha$ takes a generic value. As we have chosen the sites 1 and 3 to be antiparallel, we 
can consider only the symmetry operations of $C_{4v}=\left\{ (), (1234), (13)(24), (1432), (13), (12)(34), (24), (14)(23) \right\}$ 
the symmetry group of the tube for $J_d>0$.
\footnote{In general, any permutation of the symmetric group $S_N$ is written as some product of cyclic permutations. 
We represent such cyclic permutations as 
\begin{eqnarray*}
(jkl \cdots m): \vec{S}_j \rightarrow \vec{S}_k, \vec{S}_k \rightarrow \vec{S}_l, \cdots, \vec{S}_m \rightarrow \vec{S}_j. 
\end{eqnarray*}
We also denote an identity operation as $()$. 
} The states are invariant under the operation $(13)(24)$.
The reflections $(13)$ or $(24)$ connect the states $\alpha$ and $\pi+\alpha$ on one side and  $\pi-\alpha$ and $2\pi-\alpha$ on the other side. The states $\alpha$ and $\pi-\alpha$ are 
related by cyclic permutations $(1234)$ and $(1432)$, and same for states $\pi+\alpha$ and $2\pi-\alpha$. The
reflections $(12)(34)$ and $(14)(23)$ transform the state $\alpha$ into the state with $2\pi-\alpha$ and $\pi-\alpha$ into $\pi+\alpha$.

%%%%%%%%%%%%%%%%%%%%%%%%%%%%%%%%%%%%%%%%%%%%%%%%%%%%%%%%%%%%%%%%%%%%%%%%%%%%%%%%%%%%%%%%%%
%%%%%%%%%%%%%%%%%%%%%%%%%%%%%%%%%%%%%%%%%%%%%%%%%%%%%%%%%%%%%%%%%%%%%%%%%%%%%%%%%%%%%%%%%%

\subsection{Low-energy effective action}\label{sec:PI_action}

We plug this ground-state solution in the expressions (\ref{eq:PI_spin_operators}) and cast these expressions in the action. Up to the second order 
in the fields, we obtain in the continuum limit the following action, 
\begin{equation}
\begin{split}
S&=\int d\tau dx \sum_{j}\bigg\{\frac{aJ_\parallel}{2} \left(S^2-m^2\right)\left(\partial_x\varphi_{j}\right)^2 \\
&+ a\left(2J_{\parallel}+\frac{J_{\perp}+J_{d}}{2}\frac{S^2}{S^2-m^2}\right)\Pi_{j}^2 \\
&+\frac{J_{\perp}}{2}\sin(\alpha)\frac{S^2-m^2}{a}(-1)^{j}\left(\varphi_{j}-\varphi_{j+1}\right)^2 \\
&+\frac{J_{\perp}+J_{d}}{4}\frac{S^2-m^2}{a}\left(\varphi_{j}-\varphi_{j+2}\right)^2\\
&+aJ_{\perp}\left(1+(-1)^{j+1}\sin(\alpha)\frac{m^2}{S^2-m^2}\right)\left(\Pi_{j}\Pi_{j+1}\right) \\
&+a\frac{J_{\perp}+J_{d}}{2}\left(1-\frac{m^2}{S^2-m^2}\right)\left(\Pi_{j}\Pi_{j+2}\right) \\
&+aJ_{\perp}\sin(\alpha)m\varphi_{j}\left(\Pi_{j-1}-\Pi_{j+1}\right) \\ 
&+i\left(\frac{S-m}{a}\right)\partial_{\tau}\varphi_{j}-i\Pi_{j}\partial_{\tau}\varphi_{j} \bigg\},
\end{split}
\label{eq:PI_action1}
\end{equation}
where $a$ denotes the lattice constant. The last two imaginary terms
come from the Berry phase part of the action. We now diagonalize the momentum part with the transformation $\vec{\Omega}=\mathrm{P}\vec{\Pi}$,
where
\begin{equation}
\mathrm{P}= \frac{1}{2} \left( \begin{array}{cccc} -1 & -1 & 1 & 1 \\ -1 & 1 & -1 & 1 \\ 1 & -1 & -1 & 1 \\ 1 & 1 & 1 & 1 \end{array} \right). 
\label{eq:PI_Pi_transformation}
\end{equation}
After applying the same transformation to the $\varphi_j$ fields, $\vec{\phi}=\mathrm{P}\vec{\varphi}$, we obtain
\begin{equation}
\begin{split}
S&=\int d\tau dx \bigg\{ \sum_j \bigg[ \frac{1}{2}\lambda_j \Omega_j^2 + \frac{1}{2} \lambda_x \left(\partial_x\phi_{j}\right)^2 \bigg]\\
&+\frac{1}{2}m_1^2 \phi_1^2 + \frac{1}{2}m_3^2 \phi_3^2 +\mu \left[ \Omega_1\phi_3 - \Omega_3\phi_1 \right]\\
&+i 2\frac{S-m}{a}\partial_{\tau}\phi_4 - i \sum_j \Omega_j \partial_{\tau} \phi_j  \bigg\},
\end{split}
\label{eq:PI_action2}
\end{equation}
where the coefficients are given by 
\begin{equation} \label{eq:action_coefficients}
\begin{split}
\lambda_{1,3} &= 4aJ_{\parallel}+2a\left[J_d + J_{\perp}(1 \pm \sin(\alpha))\right] \frac{m^2}{S^2-m^2}, \\
\lambda_2 &= 4aJ_{\parallel}+2aJ_d, \\
\lambda_4 &= 4aJ_{\parallel}+2a\left(J_d+2J_{\perp}\right), \\
\lambda_x &= a J_{\parallel}\left(S^2-m^2\right), \\
m_{1,3}^2 &= 2\frac{S^2-m^2}{a}\left[J_d + J_{\perp}(1 \pm \sin(\alpha))\right], \\
\mu &= 2mJ_{\perp}\sin(\alpha). 
\end{split}
\end{equation}
Finally we can 
integrate out the massive fields $\Omega_j$ and the action reads
\begin{equation}
\begin{split}
S&=\int d\tau dx \bigg\{ \sum_j \bigg[ \frac{1}{2\lambda_j} \left(\partial_{\tau}\phi_{j}\right)^2  + \frac{1}{2\lambda_x} \left(\partial_x\phi_{j}\right)^2 \bigg] \\
&+ \frac{1}{2}\left(m_1^2 -\frac{\mu^2}{\lambda_3}\right) \phi_1^2 + \frac{1}{2}\left(m_3^2 -\frac{\mu^2}{\lambda_1} \right)\phi_3^2 \\
&+i \mu \left( \frac{1}{\lambda_1}\phi_3 \partial_{\tau}\phi_1 - \frac{1}{\lambda_3}\phi_1 \partial_{\tau}\phi_3  \right) + i 2 \frac{S-m}{a} \partial_{\tau}\phi_4 \bigg\}.
\end{split}
\label{eq:PI_action3}
\end{equation}

An important comment is to be made here about the form of the action for the $\phi_2$ field. We want to stress the absence of a mass term $m_2^2\phi_2^2$ and 
that we simply end up with a free field action. Coming back to the original $\varphi_j$ variables, we see that this $\phi_2$ field corresponds to moving together 
spins 1 and 3 on one hand and spins 2 and 4 on the other hand. We recover the fact 
that classically this deformation has no energy cost. However, as pointed out previously, 
this free angle does not arise from the symmetry of the Hamiltonian. The $\mathrm{U}(1)$ symmetry is encoded in the symmetric $\phi_4$ field, 
thus we do not expect this action to reflect the true behavior of the $\phi_2$ field. At higher ordres, a localizing potential is thus required such that the unphysical free-field nature $\phi_2$ appearing in the action is removed. Its shape, or more precisely its number of minima, is given by the form of the free energy as a function of $\alpha$, with two or four minima (see the discussion of the classical ground state).

In addition, inspecting Eq.~\eqref{eq:PI_action3}, we point out that some values of $\alpha$ play a particular role. We remark that, if $J_d=0$, for $\alpha=0$ (resp. $\pi$), the two fields $\phi_1$ and
$\phi_3$ decouples as $\mu=0$ while at the same time the mass term $m_3^2-\mu^2/\lambda_1^2$ ($m_1^2-\mu^2/\lambda_3^2$) vanishes. Thus we end with the field $\phi_3$ ($\phi_1$) to be also massless while the
other $\phi_1$ $(\phi_3)$ retains a mass term. The explanation is the same
than for the $\phi_2$ field because, when $J_d=0$ and only in that case, we can make a deformation with no energy cost by pairing spins 1 and 4 (1 and 2) and spins 2 and 3 (3 and 4). Following exactly the same reasoning as above, we expect a localizing potential at higher orders.

Another couple of special points is  $\alpha=\pi/2,3\pi/2$. Indeed in that case all the coefficients of the fields $\phi_1$ and $\phi_3$
are equal. We can then, as for the three-leg spin tube model~\cite{Plat2012}, introduce two conjugate fields $\Psi=\phi_1+i\phi_3$ and $\Psi^*=\phi_1-i\phi_3$, which represent the chirality degrees
of freedom. Despite the presence of a mass term $M^2|\Psi|^2$, the imaginary-time derivative term has been shown to have strong effects, 
and in particular to allow the possible appearance of gapless phases for $\Psi$. However, we will see below that it does not happen in the present system as those 
values are not favored by the fluctuation.

%%%%%%%%%%%%%%%%%%%%%%%%%%%%%%%%%%%%%%%%%%%%%%%%%%%%%%%%%%%%%%%%%%%%%%%%%%%%%%%%%%%%%%%%%%
%%%%%%%%%%%%%%%%%%%%%%%%%%%%%%%%%%%%%%%%%%%%%%%%%%%%%%%%%%%%%%%%%%%%%%%%%%%%%%%%%%%%%%%%%%

\subsection{Free energy and ground-state selection}\label{sec:PI_gs_selection}

In this section, we now compute the free energy and minimize it with respect to $\alpha$ to see which value is selected by the quantum fluctuation. 
We will also consider the classical limit to investigate the effect of the thermal fluctuation. 
From Eq.~\eqref{eq:PI_action3}, the action can be separated into two pieces. One contains the coupled fields $\phi_1$ and $\phi_3$ 
with coefficients depending on the angle $\alpha$, and another part is independent on $\alpha$ for $\phi_2$ and $\phi_4$. 
In the following, we are interested only in the $\alpha$-dependent part, thus from now on we drop the part for fields $\phi_2$ and $\phi_4$.

We rewrite the action by the Fourier transformation and we obtain
\begin{equation}
\begin{split}
&S = \frac{1}{2} \sum_{k,\omega_n} \bigg\{ \left[ \frac{1}{\lambda_1} \omega_n^2 + \frac{1}{\lambda_x} k^2 + \left(m_1^2 -\frac{\mu^2}{\lambda_3}\right)\right] |\phi_1(k,\omega_n)|^2  \\
&+ \left[ \frac{1}{\lambda_3} \omega_n^2 + \frac{1}{\lambda_x} k^2 + \left(m_3^2 -\frac{\mu^2}{\lambda_1}\right)\right] |\phi_3(k,\omega_n)|^2 \\
& + 2\mu \omega_n \left[ \frac{1}{\lambda_3} \phi_1(k,\omega_n) \phi_3^*(k,\omega_n) -  \frac{1}{\lambda_1} \phi_1^*(k,\omega_n) \phi_3(k,\omega_n) \right] \bigg\}\\
&= \frac{1}{2} \sum_{k,\omega_n} 
\begin{pmatrix}
\phi_1^* \\
\phi_3^* \\
\end{pmatrix}^T \mathcal{M} 
\begin{pmatrix}
\phi_1 \\
\phi_3 \\
\end{pmatrix}, 
\end{split}
\label{eq:PI_action4}
\end{equation}
where the $\omega_n=2\pi n / \beta, n\in\mathbb{Z}$ ($\beta$ being the inverse temperature) are the bosonic Matsubara frequencies and we have used the definition, 
\begin{equation}
\phi_j(x,\tau) = \frac{1}{\sqrt{\beta L}} \sum_{k,\omega_n} \mathrm{e}^{i(kx-\omega_n\tau)}\phi_j(k,\omega_n), 
\label{eq:PI_fourier_def}
\end{equation}
for the Fourier transformation. We can evaluate the partition function $\mathcal{Z}=\mathrm{Tr} e^{-S}$ and we find
\begin{equation}
\mathrm{ln}(\mathcal{Z})= \mathrm{N}(\beta) - \frac{1}{2} \sum_{k,\omega_n} \mathrm{ln}(\mathrm{det}\mathcal{M}), 
\label{eq:PI_free_energy1}
\end{equation}
up to an additional unimportant constant. The $\mathrm{N}(\beta)$ term comes from the previous integration of the $\Omega_j$ fields.~\cite{Bernard1974}  
After some manipulations, we can write
\begin{equation}
\begin{split}
\mathrm{ln}(\mathcal{Z}) &= \mathrm{N}'(\beta) - \frac{1}{2}  \sum_{k,\omega_n}  \mathrm{ln}(\omega_n^4+p\omega_n^2+q)    \\
&= \mathrm{N}'(\beta) - \frac{1}{2}  \sum_{k,\omega_n} \left[ \mathrm{ln}(\omega_n^2+\omega_+^2) + \mathrm{ln}(\omega_n^2+\omega_-^2)  \right],
\end{split}
\label{eq:PI_free_energy2}
\end{equation}
where $\omega_{\pm}^2 = (p \pm \sqrt{\Delta})/2$, $\Delta = p^2-4q$, and $p,q$ are complicated functions of the coefficients in the action (\ref{eq:PI_action4}) and 
contain the $\alpha$-dependence of the partition function. Finally, we perform the sum on the Mastubara frequencies and obtain the standard expression for the free 
energy $\mathcal{F}=-\mathrm{ln}(\mathcal{Z})/\beta$, 
\begin{equation}
\mathcal{F} = \sum_k \left\{ \frac{\omega_+ + \omega_-}{2} + \frac{1}{\beta}  \mathrm{ln}\left( \left[1-e^{-\beta\omega_+}\right] \left[1-e^{-\beta\omega_-}\right] \right)  \right\}.
\label{eq:PI_free_energy3}
\end{equation}
The $\mathrm{N}'(\beta)$ has been canceled during the summation of the $\omega_n$ frequencies.~\cite{Bernard1974} The first term is the zero-point energy and 
represents the effect of the quantum fluctuation, while the second term, vanishing in the limit of large $\beta$, corresponds to the thermal 
fluctuation. Using this expression, we now evaluate numerically the sum over momentum and minimize it with respect to $\alpha$.

\begin{figure}
\includegraphics[width=0.46\textwidth,clip]{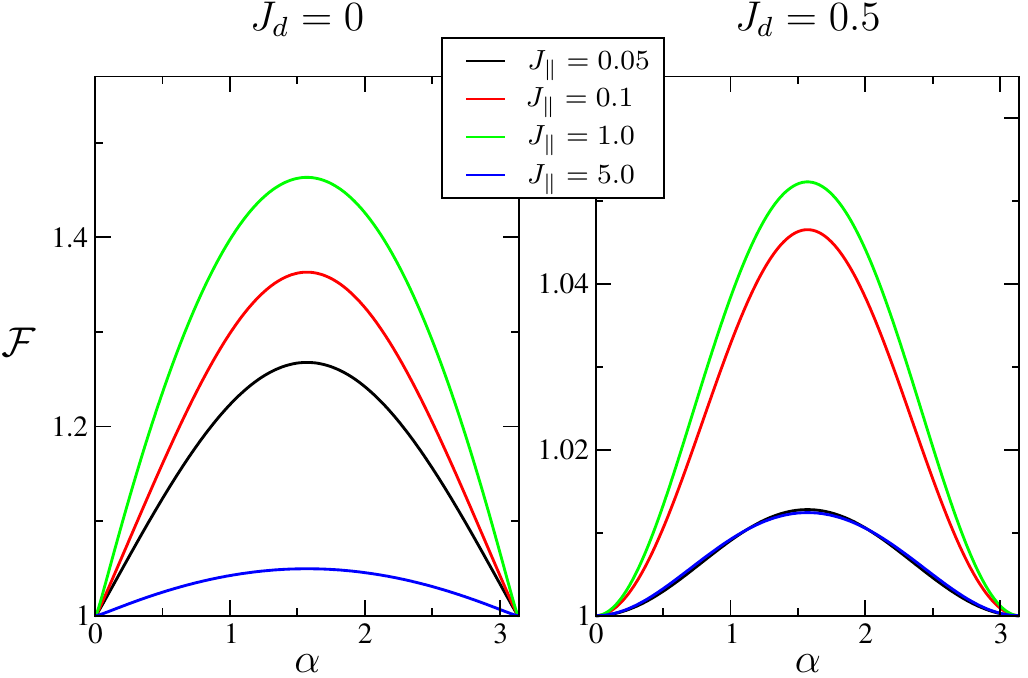}
\caption{(Color online) Numerical calculation of the free energy from Eq.~(\ref{eq:PI_free_energy3}), normalized by $\mathcal{F}(\alpha^*)$, for $S=1/2$ and $m=1/4$ 
in the classical limit. We display the cases for $J_d=0$ (left panel) and $J_d=0.5$ (right panel).}
\label{fig:free_energy_Cl}
\end{figure}

We first begin by examining the effect of the thermal fluctuation by taking the classical limit. In Fig.~\ref{fig:free_energy_Cl}, we show the free energy calculated in the classical regime
for different values of $J_{\parallel}$ and $J_{d}>0$, in unit of $J_{\perp}$. We see that the minima are always located at the colinear configurations $\alpha^*=0$ and $\pi$ 
for all the coupling values.

\begin{figure}
\includegraphics[width=0.46\textwidth,clip]{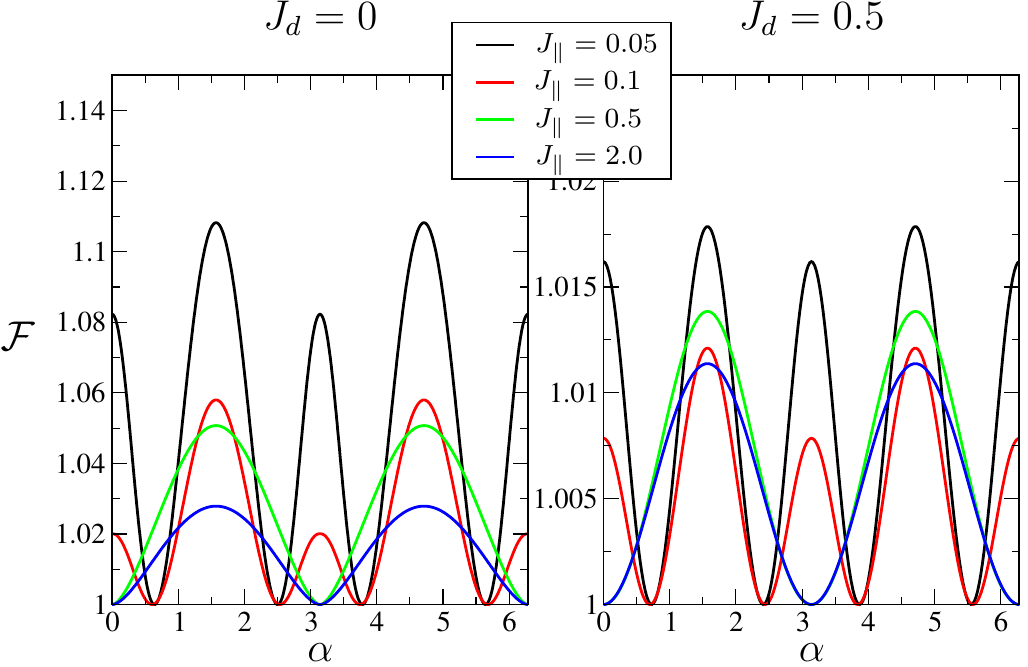}
\caption{(Color online) Numerical calculation of the free energy from Eq.~(\ref{eq:PI_free_energy3}), normalized by $\mathcal{F}(\alpha^*)$, for $S=1/2$ and $m=1/4$ 
in the quantum limit $\beta\to\infty$. We display the cases for $J_d=0$ (left panel) and $J_d=0.5$ (right panel).}
\label{fig:free_energy_Q}
\end{figure}

The quantum limit $\beta\to\infty$ where only the zero-point energy contributes is more interesting. 
We plot in Fig.~\ref{fig:free_energy_Q} the free energy as a function of $\alpha$ for several values of $J_{\parallel}$, with $S=1/2$ and a magnetization $m=1/4$, 
corresponding to a possible magnetization plateau in the quantum system from the Oshikawa-Yamanaka-Affleck condition~\cite{Oshikawa1997} $4(S-m)\in\mathbb{Z}$.
For both $J_d=0$ and $J_d>0$, we observe two regimes. First, at large $J_{\parallel}$ we find the same 
behavior that for the thermal fluctuation with two minima at $\alpha^*=0,\pi$. But, for small values of 
$J_{\parallel}$, the free energy is minimized at nontrivial values of $\alpha$, so that we get four minima at $\alpha^*$, $\pi-\alpha^*$, $\pi+\alpha^*$, and $2\pi-\alpha^*$, as discussed in 
Sec.~\ref{sec:PI_classical_gs}. Notice also the form of the free energy showing that there are two groups of minima, because of the presence of two different energy barriers. Indeed,
a large barrier at $\alpha=\pi/2$ separates the two minima at $\pi-\alpha^*$ and $\pi+\alpha^*$ from the two others, while the separation between them at $\alpha=\pi$ is smaller.

It is surprising at first sight that the two types of
fluctuation act in different directions, contrary to the case of the $J_1-J_2$ XY model on the square lattice for example.~\cite{Henley1989} 
The thermal and quantum fluctuations play the same role in most cases as we said before that for $J_d=0$, selecting $\alpha^{*}=0$ or $\pi$ implies having another field whose mass vanishes, and such a state should be favored by the fluctuations in the usual picture of OBD. However, it is important to note that the zero-point energy depends on the sum $\omega_+ + \omega_-$, whereas the thermal part is basically determined by the product $\omega_+\omega_-$ (expand the second term in Eq.~\eqref{eq:PI_free_energy3} when $\beta\to0$). 
Thus the two fluctuations can in principle have distinct effects~\cite{Cabra2005} and select different states. It
would be interesting to find a 2D or 3D system exhibiting this property as it would induce a phase transition when lowering the temperature.

%%%%%%%%%%%%%%%%%%%%%%%%%%%%%%%%%%%%%%%%%%%%%%%%%%%%%%%%%%%%%%%%%%%%%%%%%%%%%%%%%%%%%%%%%%
%%%%%%%%%%%%%%%%%%%%%%%%%%%%%%%%%%%%%%%%%%%%%%%%%%%%%%%%%%%%%%%%%%%%%%%%%%%%%%%%%%%%%%%%%%

\subsection{Discussion}\label{sec:PI_discussion}

Beyond the question of the value of the selected angles, we have previously discussed 
the fact that at this selection is associated a localizing potential for the $\phi_2$ field in Eq.~(\ref{eq:PI_action3}). 
This raises the question of the possible tunneling between the different minima.~\cite{Chandra1990}

Let us start with the simplest case of 
the regime of large $J_{\parallel}$ corresponding to $\alpha^*=0,\pi$. The double-well form of the free energy implies the emergence of a $\mathbb{Z}_2$ symmetry and two scenarios are then possible. 
If the energy barrier between the two minima is sufficiently small and that at the same time the stiffness of the $\phi_2$ field (given by $1/\lambda_2$ at this gaussian order) is also small enough, the tunneling between the minima of the potential becomes relevant and therefore the emergent $\mathbb{Z}_2$ symmetry is unbroken. This corresponds to a unique ground state for both $J_d=0$ and $J_d>0$. On the opposite limit of a energy barrier too large compare to the field stiffness,
the tunneling between the minima is suppressed and the $\mathbb{Z}_2$ symmetry is broken. In that case, for $J_d>0$ we expect the ground state to be two-fold degenerate and for $J_d=0$ three-fold 
degenerate, because we could have started the calculation from a classical configuration with spin 2 or 4 antiparallel with spin 1, and this adds one more distinct state (remind the initially broken $\mathbb{Z}_3$ symmetry). This state corresponds to a $k_{\parallel}=0$ ordering of the operator $Q^{1234}$ since it takes different values at the two minima 
(the third state also takes a different one), and the chirality has a zero expectation
value. It is however important to remind that we are working with a one-dimensional system. Thus 
in the case of thermal fluctuation, thermal activation is always possible since the $\mathbb{Z}_2$ symmetry
cannot be broken at finite temperature, and only in the quantum case the above discussion is relevant. From our calculation at the gaussian order, it is however not possible to give quantitative predictions about wether this $\mathbb{Z}_2$ symmetry is broken as we do not have access to the value of the potential.

We consider now the case of $\alpha^*\neq 0$ or $\pi$ that we found in the regime of small $J_{\parallel}$. Given the existence of four minima and two different energy barriers,
the situation is more complex and two kinds of tunneling have to be considered. However, we will see that, in this regime of moderately small $J_{\parallel}$, the situation is actually more complicated 
as a $k_{\parallel}=\pi$ ordering appears. In the expansion (\ref{eq:PI_spin_operators}), by keeping the same unit cell of four spins, we assumed that any ordering would be at $k_{\parallel}=0$, thus 
such a phase cannot be described in
our calculation. It would require the addition of more degrees of freedom by doubling of the unit cell and working with eight fluctuation fields, which will be discussed in Sec.~\ref{sec:comp_path_int}.

We also want to show that the relevance of the tunneling opens the possibility of observing {\it quantized} spin imbalance phases. 
To begin with, using the relation $S^z_j=m+\Pi_j$, we see that the way to obtain a different magnetization depending on the chain is to
have a nonzero value for one or several of the fields $\Omega_j$. In the following we are interested in the $\Omega_2$ field, for which the corresponding spin imbalance pattern 
is, like for $\phi_2$, grouping spins 1 and 3 on one side and spins 2 and 4 on the other side such that $\langle S^z_1-S^z_2+S^z_3-S^z_4\rangle = 2\langle\Omega_2\rangle \neq 0$. Thus, as the effective potential becomes sufficiently flat together with a value of the stiffness favoring the tunneling, the field becomes more and more delocalized, i.e. $\Delta\phi_2$ becomes very large. As a consequence the wave function
gets closer to a plane wave. The key point is then to notice that the field $\Omega_2$ is thus strongly locked to its eigenvalues due to the uncertainty principle ($\Delta\Omega_2\to 0$),
since the original $\Pi_j$ variables have been defined to be the conjugate momentum to the angular fluctuations 
$\varphi_j$. Because these variables are defined between 0 
and $2\pi$, the $\Pi_j$ have integer eigenvalues $0,\pm1,\pm2,\cdots$ and this translates into half-integer eigenvalues for the fields 
$\Omega_j$ according to transformation (\ref{eq:PI_Pi_transformation}). Then a spin imbalance phase
associated to $\Omega_2$ would be automatically quantized to an integer value, namely $\langle S^z_1-S^z_2+S^z_3-S^z_4\rangle = 0,\pm1,\pm2, \cdots $. Obviously, because of the
eigenvalue 0 it is also possible to get no spin imbalance, and this is what the action (\ref{eq:PI_action2}) would predict with only the kinetic term $\Omega_2^2$. But even in this case,
the locking mechanism would manifest itself by strongly suppressing the fluctuation of the spin imbalance observable. 
It is worth reminding that in our analysis the spin imbalance is predicted to be a uniform $k_{\parallel}=0$ phase. We will elaborate on those two points after reporting the strong-coupling
and numerical results where we obtain a {\it staggered}, thus $k_\parallel = \pi$, quantized spin imbalance phase.

We want to emphasize the specificity of such a spin imbalance phase, whose nature is very distinct
from the spin imbalance phases reported in the Heisenberg model in a magnetic field on two different three-leg spin tubes (one uniform phase and one staggered). In both cases, the spin imbalance
magnitude is not constrained to take any specific value and varies with the longitudinal spin coupling $J_{\parallel}$.~\cite{Okunishi2012, Plat2012} 
Here, the locking to quantized values also tells us that the order parameter measuring the spin imbalance is basically insensitive to the Hamiltonian parameters. This difference stems directly
from the continuous degeneracy of the classical ground state and the following OBD effect present in this model while absent for the three-leg tube. We will show
analytical results from perturbation theory and numerical simulations confirming this robustness.

%%%%%%%%%%%%%%%%%%%%%%%%%%%%%%%%%%%%%%%%%%%%%%%%%%%%%%%%%%%%%%%%%%%%%%%%%%%%%%%%%%%%%%%%%%
%%%%%%%%%%%%%%%%%%%%%%%%%%%%%%%%%%%%%%%%%%%%%%%%%%%%%%%%%%%%%%%%%%%%%%%%%%%%%%%%%%%%%%%%%%
%%%%%%%%%%%%%%%%%%%%%%%%%%%%%%%%%%%%%%%%%%%%%%%%%%%%%%%%%%%%%%%%%%%%%%%%%%%%%%%%%%%%%%%%%%

\section{Strong-coupling expansion}\label{sec:Strong_coupling}

We present a strong-coupling analysis of the model (\ref{eq:4leg_tube_Ham}) by deriving effective Hamiltonians up to the second order in the
coupling $J_{\parallel}$. We first analyze the $S=1/2$ case on the magnetization plateau $m=1/4$ and show the appearance of spin imbalance phases.
Then, we move to the general spin-$S$ for which a new phase appears, and we investigate the nature of the phase transition coming from the spin imbalance regime.

%%%%%%%%%%%%%%%%%%%%%%%%%%%%%%%%%%%%%%%%%%%%%%%%%%%%%%%%%%%%%%%%%%%%%%%%%%%%%%%%%%%%%%%%%%
%%%%%%%%%%%%%%%%%%%%%%%%%%%%%%%%%%%%%%%%%%%%%%%%%%%%%%%%%%%%%%%%%%%%%%%%%%%%%%%%%%%%%%%%%%

\subsection{Single tetrahedron for \texorpdfstring{$S=1/2$}{S=1/2}}

We consider here a single tetrahedron of $S=1/2$ spins with the Hamiltonian, 
\begin{eqnarray}
H_0 = J_\perp \sum_{j<j'} \vec{S}_j \cdot \vec{S}_{j'} + J_d \left( \vec{S}_1 \cdot \vec{S}_3 + \vec{S}_2 \cdot \vec{S}_4 \right) 
   -h \sum_{i=1}^4 S^z_j. \nonumber \\
\end{eqnarray}
At $J_d=0$, this Hamiltonian has an $S_4$ symmetry (or equivalently, a tetrahedral $T_d$ symmetry) corresponding to any permutation of the four spins. 
We note that the $S_4$ symmetry can be decomposed into its subgroups, such as $S_4 = \mathbb{Z}_4 \times \mathbb{Z}_3 \times \mathbb{Z}_2$, 
where $\mathbb{Z}_4=\{ (),(1234),(13)(24),(1432) \}$, $\mathbb{Z}_3=\{ (),(123),(132) \}$, and $\mathbb{Z}_2=\{ (),(13) \}$.
This decomposition is useful to understand the symmetry properties of eigenstates of a single tetrahedron and the effective Hamiltonians in the following discussion.

\begin{figure}
\includegraphics[width=0.35\textwidth,clip]{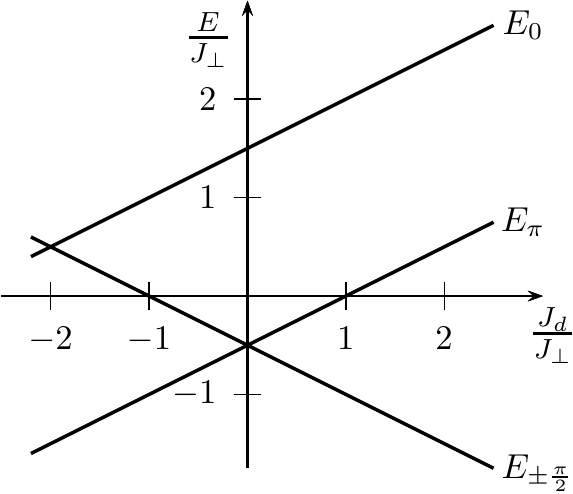}
\caption{Plot of the eigenvalues of a single tetrahedron as a function of $J_d/J_{\perp}$.}
\label{fig:eigenvalues_m1_4}
\end{figure}

If introducing the diagonal asymmetry $J_d \neq 0$, the $S_4$ symmetry breaks down to a $C_{4v}=\mathbb{Z}_4 \times \mathbb{Z}_2$ symmetry. 
Thus we can choose eigenstates of $H_0$ as ``momentum'' eigenstates $\left| k_\square \right>$ 
to satisfy $P_{\square} \left| k_\square \right> = k_\square \left| k_\square \right>$, 
which respects the $\mathbb{Z}_4$ symmetry corresponding to the cyclic permutation of four spins, $P_\square$: $\vec{S}_j \rightarrow \vec{S}_{j+1}$. 
Then the four eigenstates with $S^z_{\boxtimes}=1$ are written as 
\begin{eqnarray}
\left| k_\square=0 \right> = \frac{1}{2} \left( \left| \da \ua \ua \ua \right> +\left| \ua \da \ua \ua \right> 
   +\left| \ua \ua \da \ua \right> +\left| \ua \ua \ua \da \right> \right) 
\end{eqnarray}
for $S_\boxtimes=2$, and 
\begin{equation}
\begin{split}
\left| \pi/2 \right> =& \frac{1}{2} \left( \left| \da \ua \ua \ua \right> +\omega \left| \ua \da \ua \ua \right> 
   +\omega^2 \left| \ua \ua \da \ua \right> +\omega^3 \left| \ua \ua \ua \da \right> \right), \\
\left| \pi \right> =& \frac{1}{2} \left( \left| \da \ua \ua \ua \right> -\left| \ua \da \ua \ua \right> 
   +\left| \ua \ua \da \ua \right> -\left| \ua \ua \ua \da \right> \right), \\
\left| -\pi/2 \right> =& \frac{1}{2} \left( \left| \da \ua \ua \ua \right> +\omega^3 \left| \ua \da \ua \ua \right> 
   +\omega^2 \left| \ua \ua \da \ua \right> +\omega \left| \ua \ua \ua \da \right> \right) \\
\end{split}
\label{eq:eigenstates_chi}
\end{equation}
for $S_\boxtimes=1$, where $\omega = \exp (i\pi/2)$ and we denote the basis vectors as $\left| S^z_1 S^z_2 S^z_3 S^z_4 \right>$. 
The corresponding energy eigenvalues are given by $E_{k_\square=0}=(3J_\perp+J_d)/2$, $E_{\pm\pi/2}=(-J_\perp-J_d)/2$, and $E_{\pi}=(-J_\perp+J_d)/2$, 
and shown in Fig.~\ref{fig:eigenvalues_m1_4} as functions of $J_d$. 
Thus we have three regimes; (i) for $J_d <0$, the ground state is in the $k_\square=\pi$ state and unique, 
(ii) for $J_d > 0$, the ground state is two-fold degenerate with a doublet of states of momentum $k_{\square}=\pm\pi/2$, 
and (iii) at $J_d =0$, these states form a three-fold degenerate ground state since $H_0$ is simply written in terms of ${\vec S}_\boxtimes$. 
The state with $k_\square =0$ is always a high energy state and is neglected in our analysis. 

\begin{figure}
\includegraphics[width=0.4\textwidth,clip]{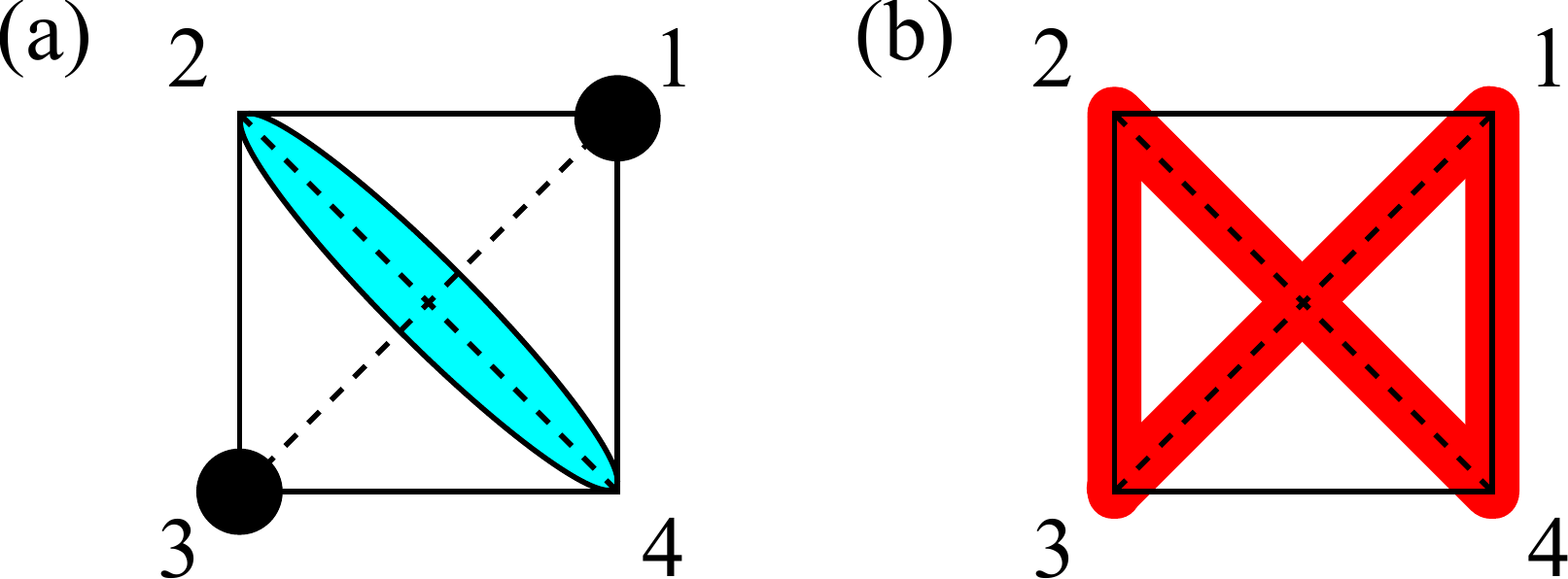}
\caption{(Color online) Schematic picture of the states (a) $\left| \Psi_{24} \right>$ in Eq.~\eqref{eq:Psi_singlets} and (b) $\left| + \right>$ in Eq.~\eqref{eq:Q_Psi}. 
A blue open circle and a filled black circle denote the singlet bond and the polarized spin, respectively. 
The links on which a singlet resonates are represented by a red thick line. }
\label{fig:Ising_phase_cartoon}
\end{figure}

On the other hand, we can also write the eigenstates of $H_0$ as the eigenstates of an operator $Q^{1324}$ defined by 
\begin{eqnarray} \label{eq:op_Q}
Q^{jklm} \equiv (\vec{S}_j \times \vec{S}_k) \cdot (\vec{S}_l \times \vec{S}_m).
\end{eqnarray} 
Recall that we also had introduced this operator in the semi-classical approach to distinguish the various classical states (Sec.~\ref{sec:PI_classical_gs}). 
This operator is symmetric under $D_2 = \{ (), (12)(34), (13)(24), (14)(23) \}$.  
The three low-energy states with $S_\boxtimes=1$ have $4Q^{1324}=1$, $-1$, and $0$, and corresponding eigenstates are given by 
\begin{equation}
\begin{split}
\left| + \right> =& \frac{1}{2} \left( \left| \da \ua \ua \ua \right> +\left| \ua \da \ua \ua \right> 
   -\left| \ua \ua \da \ua \right> -\left| \ua \ua \ua \da \right> \right), \\
\left| - \right> =-& \frac{1}{2} \left( \left| \da \ua \ua \ua \right> -\left| \ua \da \ua \ua \right> 
   -\left| \ua \ua \da \ua \right> +\left| \ua \ua \ua \da \right> \right), \\
\left| 0 \right> =& \frac{1}{2} \left( \left| \da \ua \ua \ua \right> -\left| \ua \da \ua \ua \right> 
   +\left| \ua \ua \da \ua \right> -\left| \ua \ua \ua \da \right> \right), 
\end{split}
\label{eq:eigenstates_X}
\end{equation}
where their energy eigenvalues are $E_\pm=(-J_\perp-J_d)/2$ and $E_0=(-J_\perp+J_d)/2$. 
These states are interpreted as linear combinations of the wave function consisting of one singlet and two polarized spins, 
\begin{eqnarray}\label{eq:Psi_singlets}
\left| \Psi_{jk} \right> = \frac{1}{\sqrt{2}} \left( \left| \ua_j \da_k \right>-\left| \da_j \ua_k \right> \right) 
   \otimes \left| \ua_l \ua_m \right>, 
\end{eqnarray}
where $l$ and $m$ represent positions of the other spins than $j$ and $k$ (see Fig.~\ref{fig:Ising_phase_cartoon}~(a)). 
Using this wave function, we can rewrite Eq.~\eqref{eq:eigenstates_X} as 
\begin{equation} \label{eq:Q_Psi}
\begin{split}
\left| + \right> =& \frac{1}{\sqrt{2}} \left( \left| \Psi_{13} \right> + \left| \Psi_{24} \right> \right) 
   \hspace{5pt} \textrm{or} \hspace{5pt} \frac{1}{\sqrt{2}} \left( \left| \Psi_{14} \right> + \left| \Psi_{23} \right> \right), \\
\left| - \right> =& \frac{1}{\sqrt{2}} \left( \left| \Psi_{12} \right> - \left| \Psi_{34} \right> \right) 
   \hspace{5pt} \textrm{or} \hspace{5pt} \frac{1}{\sqrt{2}} \left( \left| \Psi_{13} \right> - \left| \Psi_{24} \right> \right), \\
\left| 0 \right> =& \frac{1}{\sqrt{2}} \left( \left| \Psi_{12} \right> + \left| \Psi_{34} \right> \right) 
   \hspace{5pt} \textrm{or} \hspace{5pt} \frac{1}{\sqrt{2}} \left( \left| \Psi_{14} \right> - \left| \Psi_{23} \right> \right), 
\end{split}
\end{equation}
This interpretation of the eigenstates will be convenient to analyze the ground-state properties of the coupled tetrahedra. 
Indeed, the eigenstate of $Q^{1324}$ is a ``tetramer'' state in which a singlet resonance only lives on the four bonds of a certain plaquette (see Fig.~\ref{fig:Ising_phase_cartoon}~(b)).

%###################################################################################################
%####################################################################################################

We note that two sets of the three eigenstates in Eq.~\eqref{eq:eigenstates_chi} and \eqref{eq:eigenstates_X} 
are related by a unitary transformation, $v_k = \mathcal{U} v_Q$ with
\begin{eqnarray} \label{eq:unitary_ktoq}
\mathcal{U} = \left( \begin{array}{ccc} \frac{1}{2} (1-i) & \frac{1}{2}(1+i) & 0 \\ 0 & 0 & 1 \\ \frac{1}{2}(1+i) & \frac{1}{2}(1-i) & 0 \end{array} \right), 
\end{eqnarray}
and
\begin{eqnarray} \label{eq:basis_ktoq}
v_k = \left( \begin{array}{c} \left| \pi/2 \right> \\ \left| \pi \right> \\ \left| -\pi/2 \right> \end{array} \right), \hspace{10pt} 
v_Q = \left( \begin{array}{c} \left| + \right> \\ \left| - \right> \\ \left| 0 \right> \end{array} \right). 
\end{eqnarray}
In the following, we introduce a leg exchange $J_\parallel$ between tetrahedra to form the four-leg tube \eqref{eq:4leg_tube_Ham} 
and derive an effective Hamiltonian in the strong-coupling limit $J_\parallel \ll J_\perp, J_d$. 
Hereafter, we call the basis vectors $v_k$ and $v_Q$ as the ``momentum basis'' and ``Q basis,'' respectively.

%%%%%%%%%%%%%%%%%%%%%%%%%%%%%%%%%%%%%%%%%%%%%%%%%%%%%%%%%%%%%%%%%%%%%%%%%%%%%%%%%%%%%%%%%%
%%%%%%%%%%%%%%%%%%%%%%%%%%%%%%%%%%%%%%%%%%%%%%%%%%%%%%%%%%%%%%%%%%%%%%%%%%%%%%%%%%%%%%%%%%

\subsection{Strong-coupling Hamiltonian} \label{sec:SCHam}

First we focus on the symmetric point $J_d=0$ where the Hamiltonian~\eqref{eq:4leg_tube_Ham} has an $S_4$ symmetry corresponding 
to any permutation of four legs. 
Since the ground state of a single tetrahedron is three-fold degenerate, we perform degenerate perturbation theory in the $3^L$-dimensional Hilbert space. 
In the Q basis, we find 
\begin{eqnarray} \label{eq:Heff_1st}
H_\textrm{eff}^{(1)} = \frac{J_\parallel}{4} \sum_{i=1}^L \left[ \lambda^1_i \lambda^1_{i+1} 
   +\lambda^4_i \lambda^4_{i+1} +\lambda^6_i \lambda^6_{i+1}  \right], 
\end{eqnarray}
where $\lambda^\alpha$, $\alpha=1, \cdots, 8$ are the Gell-Mann matrices (for the definition, see Appendix~\ref{sec:GellMann}).
This effective Hamiltonian obviously has a $\mathbb{Z}_3$ symmetry corresponding to the cyclic permutation of three basis vectors, 
associated with the cyclic permutation of three of four legs in the original tube, while the $\mathbb{Z}_4$ symmetry is hidden.  
$\mathbb{Z}_3$ symmetry is given by the group elements $\left\{ 1, \mathcal{X}, \mathcal{X}^2 \right\}$ with 
\begin{eqnarray} \label{eq:C3op}
\mathcal{X} = \left( \begin{array}{ccc} 0 & 1 & 0 \\ 0 & 0 & 1 \\ 1 & 0 & 0 \end{array} \right), \hspace{10pt} 
\mathcal{X}^2 = \left( \begin{array}{ccc} 0 & 0 & 1 \\ 1 & 0 & 0 \\ 0 & 1 & 0 \end{array} \right), 
\end{eqnarray}
and associated cyclic permutations of three legs are $\mathcal{X}=(132)$ and $\mathcal{X}^2=(123)$. 

As we will see in Sec.~\ref{sec:Hidden_ferro}, Eq.~\eqref{eq:Heff_1st} has a hidden SU(2) symmetry leading to a macroscopically degenerate ground state.  
To lift this massive degeneracy, it is necessary to add the second-order perturbation in $J_\parallel$. 
In the Q basis, the second-order effective Hamiltonian is given by 
\begin{widetext}
\begin{eqnarray} \label{eq:Heff_2nd}
H_\textrm{eff}^{(2)} &=& \sum_{i=1}^L \left[ 
   q_1 \left( \lambda^1_i \lambda^1_{i+1} +\lambda^4_i \lambda^4_{i+1} +\lambda^6_i \lambda^6_{i+1} \right) 
   +q_2 \left( \lambda^2_i \lambda^2_{i+1} +\lambda^5_i \lambda^5_{i+1} +\lambda^7_i \lambda^7_{i+1} \right) 
   +q_3 \left( \lambda^3_i \lambda^3_{i+1} +\lambda^8_i \lambda^8_{i+1} \right) \right. \nonumber \\
   && +t_1 \left( \lambda^1_i \lambda^4_{i+1} \lambda^6_{i+2} +\lambda^4_i \lambda^6_{i+1} \lambda^1_{i+2} +\lambda^6_i \lambda^1_{i+1} \lambda^4_{i+2}
        +\lambda^1_i \lambda^6_{i+1} \lambda^4_{i+2} +\lambda^6_i \lambda^4_{i+1} \lambda^1_{i+2} +\lambda^4_i \lambda^1_{i+1} \lambda^6_{i+2} \right) \nonumber \\
&& \left. +t_2
   \left( \lambda^1_i \left( 1 -\sqrt{3} \lambda^8_{i+1} \right) \lambda^1_{i+2} 
   + \lambda^4_i \left( 1 -\frac{3}{2} \lambda^3_{i+1} +\frac{\sqrt{3}}{2} \lambda^8_{i+1} \right) \lambda^4_{i+2} 
   + \lambda^6_i \left( 1 +\frac{3}{2} \lambda^3_{i+1} +\frac{\sqrt{3}}{2} \lambda^8_{i+1} \right) \lambda^6_{i+2} \right) \right], \nonumber \\
\end{eqnarray}
\end{widetext}
where the coupling constants are 
\begin{eqnarray}
&& q_1 = \frac{J_\parallel}{4}+\frac{7J_\parallel^2}{128J_\perp}, \hspace{10pt}
q_2 = -\frac{31J_\parallel^2}{128J_\perp}, \hspace{10pt}
q_3 = -\frac{33J_\parallel^2}{128J_\perp}, \nonumber \\
&& t_1 = -\frac{J_\parallel^2}{32J_\perp}, \hspace{10pt}
t_2 = -\frac{J_\parallel^2}{48J_\perp}. 
\end{eqnarray}

Next we consider the diagonal asymmetry $J_d$ in the four-leg tube Hamiltonian~\eqref{eq:4leg_tube_Ham}. 
This introduces a ``magnetic field'' which explicitly breaks the $\mathbb{Z}_3$ symmetry in the Q basis, and the first-order Hamiltonian is modified as 
\begin{eqnarray} \label{eq:Heff_asym}
H_\textrm{eff}^{(1)} = \sum_{i=1}^L \left[ \frac{J_\parallel}{4} \left( \lambda^1_i \lambda^1_{i+1} 
   +\lambda^4_i \lambda^4_{i+1} +\lambda^6_i \lambda^6_{i+1} \right) -\frac{J_d}{\sqrt{3}} \lambda^8_i \right], \nonumber \\
\end{eqnarray}
up to an additive constant. 
The magnetic field couples with $\lambda^8_i$ and favors one (resp. two) of the three states on each site for $J_d<0$ (resp. $J_d>0$), 
as seen from Fig.~\ref{fig:eigenvalues_m1_4}. 

%%%%%%%%%%%%%%%%%%%%%%%%%%%%%%%%%%%%%%%%%%%%%%%%%%%%%%%%%%%%%%%%%%%%%%%%%%%%%%%%%%%%%%%%%%%

\subsection{Order parameters}

%########################################################################
\begin{table*}
\caption{Several order parameters relevant in this paper. Their symmetry properties in the symmetric group (SG) and the point group (PG) languages are also displayed.}
\label{table:OrderParam}
\begin{ruledtabular}
\begin{tabular}{cccc}
Symbol & Order parameter & SG sym. & PG sym. \\ \hline
$\mu_i^{jklm}$ & $S^z_{i,j}-S^z_{i,k}+S^z_{i,l}-S^z_{i,m}$ & $\{ (),(jl),(km),(jl)(km) \}$ & $C_{2v}$ \\
$\chi_i^{jklm}$ & $(\vec{S}_{i,j} \times \vec{S}_{i,k})^z + (\vec{S}_{i,k} \times \vec{S}_{i,l})^z +(\vec{S}_{i,l} \times \vec{S}_{i,m})^z + (\vec{S}_{i,m} \times \vec{S}_{i,j})^z$ & $\{ (),(jklm),(jl)(km),(jmlk) \}$ & $S_4$ \footnote{Here $S_4$ means the rotatory reflection symmetry.} \\
$Q_i^{jklm}$ & $( \vec{S}_{i,j} \times \vec{S}_{i,k} ) \cdot ( \vec{S}_{i,l} \times \vec{S}_{i,m} )$ & $\{ (),(jk)(lm),(jl)(km),(jm)(kl) \}$ & $D_2$ \\
$P_i^{jklm}$ & $(\vec{S}_{i,j}+\vec{S}_{i,l}) \cdot (\vec{S}_{i,k}+\vec{S}_{i,m}) -2(\vec{S}_{i,j} \cdot \vec{S}_{i,k} + \vec{S}_{i,l} \cdot \vec{S}_{i,m})$ & $\{ (),(jklm),(jl)(km),(jmkl) \} \{ (),(jl) \}$ & $D_{2d}$
\end{tabular}
\end{ruledtabular}
\end{table*}
%########################################################################

We here provide the connection between physical operators in the original tube~\eqref{eq:4leg_tube_Ham} and 
the Gell-Mann matrices appearing in the effective Hamiltonian. 
In Table.~\ref{table:OrderParam}, we define several operators on rung $i$, which detects spontaneous breaking of the $S_4$ symmetry. 
$\chi^{jklm}_i$ measures the $z$ component of the spin vector chirality on the plaquette $(jklm)$. 
The momentum eigenstates defined in Eq.~\eqref{eq:eigenstates_chi} are eigenstates of the operator $\chi^{1234}_i$. 
$\mu^{jklm}_i$ measures the rung spin imbalances associated with the formation of two different dimers on the opposite bonds $(jl)$ and $(km)$ as in Fig.~\ref{fig:Ising_phase_cartoon}~(a). 
$Q^{jklm}_i$ measures the formation of two different dimers on two pairs of bonds $[(jk),(lm)]$ and $[(kl),(jm)]$, while $P^{jklm}_i$ measures the tetramer formation on the plaquette $(jklm)$ as in Fig.~\ref{fig:Ising_phase_cartoon}~(b). 
We also define a projection operator onto the subspace spanned by the three eigenstates~\eqref{eq:eigenstates_X} at rung $i$ as $\mathcal{P}_i = v_Q v_Q^\dagger$ (see Eq.~\eqref{eq:basis_ktoq}). 
The above operators are represented by the Gell-Mann matrices in the truncated space: 
\begin{equation} \label{eq:rung_ops}
\begin{split}
& \mathcal{P}_i \mu^{1234}_i \mathcal{P}_i = \lambda^1_i, \hspace{10pt} 
\mathcal{P}_i \chi^{1234}_i \mathcal{P}_i = \lambda^2_i, \\ 
& \mathcal{P}_i \mu^{3124}_i \mathcal{P}_i = \lambda^4_i, \hspace{10pt} 
\mathcal{P}_i \chi^{3124}_i \mathcal{P}_i = -\lambda^5_i, \\
& \mathcal{P}_i \mu^{2314}_i \mathcal{P}_i = \lambda^6_i, \hspace{10pt} 
\mathcal{P}_i \chi^{2314}_i \mathcal{P}_i = \lambda^7_i, \\
& \mathcal{P}_i Q^{1324}_i \mathcal{P}_i = \frac{1}{4} \lambda^3_i, \hspace{10pt} 
\mathcal{P}_i P^{1234}_i \mathcal{P}_i = \sqrt{3} \lambda^8_i. 
\end{split}
\end{equation}
One can easily see that the other $Q$ operators are obtained by the $\mathbb{Z}_3$ operations $\mathcal{X}$ and $\mathcal{X}^2$, 
\begin{equation}
\begin{split}
\mathcal{P}_i Q^{3214}_i \mathcal{P}_i &= \frac{1}{4} \mathcal{X} \lambda^3_i \mathcal{X}^{-1}, \\
\mathcal{P}_i Q^{2134}_i \mathcal{P}_i &= \frac{1}{4} \mathcal{X}^2 \lambda^3_i \mathcal{X}^{-2}, 
\end{split}
\end{equation}
and $P$ is related to $Q$ by 
\begin{eqnarray} \label{eq:PandQ}
\mathcal{P}_i P^{1234}_i \mathcal{P}_i = \mathcal{P}_i (Q_i^{2134}-Q^{3214}_i) \mathcal{P}_i. 
\end{eqnarray}

We note that $Q^{1324}_i$ and $P^{1234}_i$ form the $E$ representation of the tetrahedral symmetry group $T_d$. 
On the other hand, $\mu$'s and $\chi$'s form the $T_2$ and $T_1$ representations, respectively. 
In the discussion of the lattice distortion on the pyrochlore lattice, the $E$ representation is relevant in zero magnetic field and leads the tetragonal or orthorhombic distortion.~\cite{Yamashita2000,Tchernyshyov2002a,Tchernyshyov2002b} 
In a magnetic field, the $T_2$ representation allows the trigonal distortion and the half-magnetization plateau at the classical level.~\cite{Penc2004}
A related $\mathbb{Z}_3 \times \mathbb{Z}_2$ symmetry breaking phase is also proposed in the presence of a Dzyaloshinskii-Moriya interaction.~\cite{Canals2008}
The $T_1$ representation generally leads to some chiral ordered state as found in the pyrochlore lattice with coupled tetrahedra in the presence of a magnetic field and Dzyaloshinskii-Moriya interaction.~\cite{Kotov2005}

%%%%%%%%%%%%%%%%%%%%%%%%%%%%%%%%%%%%%%%%%%%%%%%%%%%%%%%%%%%%%%%%%%%%%%%%%%%%%%%%%%%%%%%%%%%

\subsection{Hidden ferromagnetism and ground-state selection} \label{sec:Hidden_ferro}

Actually, besides the discrete $S_4$ symmetry coming from the original spin tube, the first-order Hamiltonian \eqref{eq:Heff_1st} possesses a hidden SU(2) symmetry under the open boundary condition (OBC). 
This model can be exactly mapped onto the spin-1 Heisenberg ferromagnet, 
\begin{eqnarray}
\calV H_\textrm{eff}^{(1)} \calV^{-1} = -\frac{J_\parallel}{4} \sum_{i=1}^L \vec{T}_i \cdot \vec{T}_{i+1}, 
\end{eqnarray}
by a nonlocal unitary transformation $\calV$ introduced by Kennedy \cite{Kennedy1994} (see Appendix~\ref{sec:hidden_sym}), where $\vec{T}_i$ is a spin-1 operator. 
We therefore obtain the exact $(2L+1)$-fold degenerate ground state with ferromagnetic order. 
If we go back to the original problem by the nonlocal unitary transformation, macroscopic degeneracy of the ground state still remains but most of the ferromagnetic states will be disordered in the same manner as the Affleck-Kennedy-Lieb-Tasaki model \cite{Affleck1987,Kennedy1992} (several exceptions are shown below). 
We note that such a hidden SU(2) symmetry has also been observed \cite{Kitazawa2003} in the spin-1 XY model under the OBC 
in which case the symmetry takes the spin-1/2 representation while the spin-1 representation in our case. 
Although the SU(2) symmetry is smeared under the periodic boundary condition (PBC), we found that a ground-state degeneracy proportional to $L$ still remains. 

Once the higher-order perturbations as in Eq.~\eqref{eq:Heff_2nd} are turned on, the system starts to ``feel'' the $S_4$ anisotropy. 
Then the emergent SU(2) symmetry is reduced to $T \times D_2 \times \mathbb{Z}_3$ where $T$, $D_2$, and $\mathbb{Z}_3$ denote time reversal, dihedral group of $\pi$ rotations around spin axes, and cyclic group of permutations of spin axes. 
We expect that, among the macroscopically degenerate ferromagnetic states, some of them are selected by the $S_4$ anisotropy. 
Although a local operator generally takes some nonlocal form through a nonlocal transformation, at least to the second order, the higher-order perturbations still take local forms (see Appendix~\ref{sec:hidden_sym}). 
Thus those states can have a well-defined usual long-range order. 

\begin{figure}
\includegraphics[width=0.47\textwidth,clip]{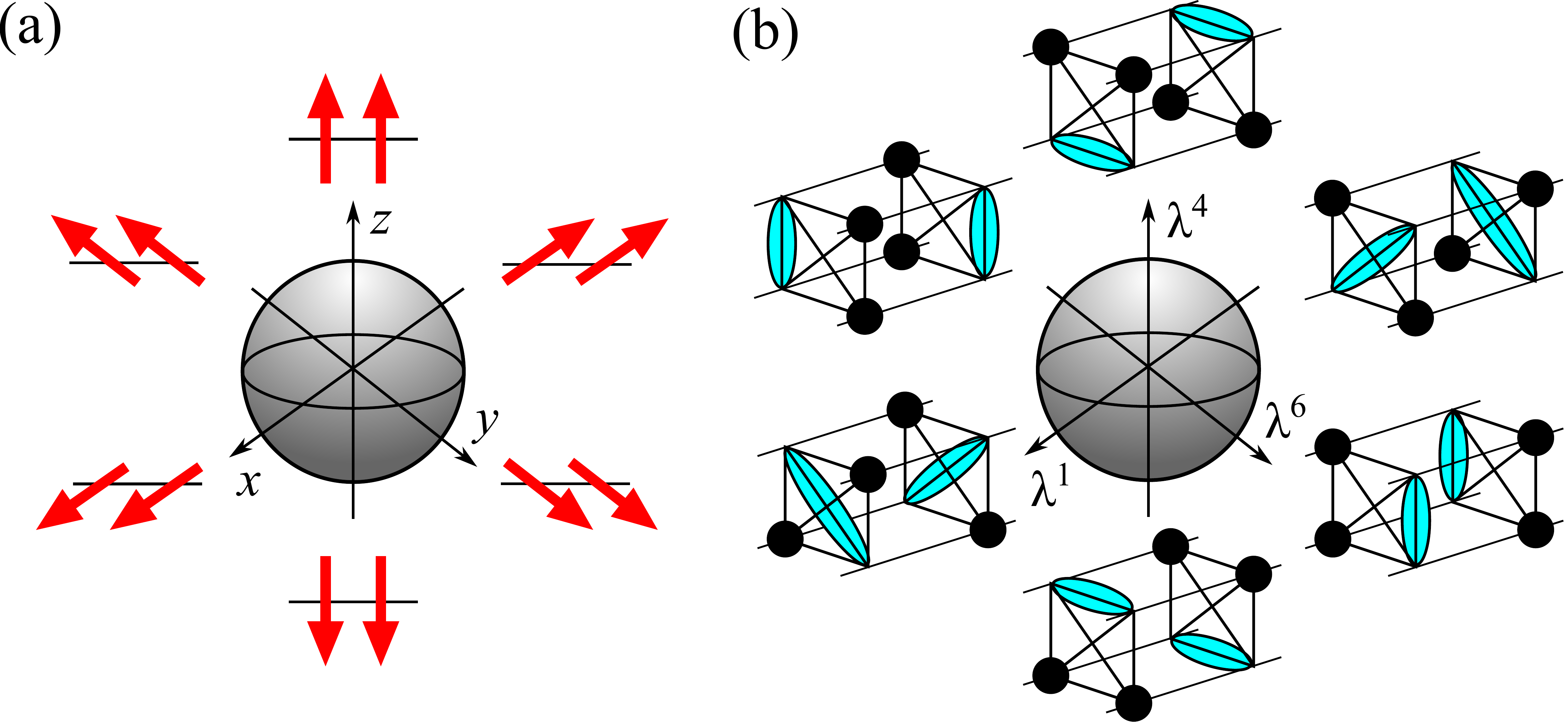}
\caption{(Color online) Schematic picture of the six-fold degenerate ground state. 
(a) In the ferromagnet after the nonlocal transformation, the ground state is a ferromagnetic state aligned in the $x$, $y$, or $z$ direction. 
(b) In the original model, the ground state is a staggered spin imbalance phase associated with $\lambda^1$, $\lambda^6$, or $\lambda^4$. 
}
\label{fig:SixGS}
\end{figure}

Indeed, we find the six-fold ferromagnetic ground state aligned in the $x$, $y$, or $z$ direction, as depicted in Fig.~\ref{fig:SixGS}~(a). 
This ferromagnetic order is related to the long-range order in the original model by the string order parameter \cite{Kennedy1994}, 
\begin{equation}
\begin{split}
\calV [(-1)^r \lambda^1_1 \lambda^1_r] \calV^{-1} &= \tilde{O}^x_\textrm{string} (r), \\
\calV [(-1)^r \lambda^4_1 \lambda^4_r] \calV^{-1} &= \tilde{O}^z_\textrm{string} (r), \\
\calV [(-1)^r \lambda^6_1 \lambda^6_r] \calV^{-1} &= \tilde{O}^y_\textrm{string} (r), 
\end{split}
\end{equation}
where
\begin{eqnarray}
\tilde{O}^\mu_\textrm{string} (r) = -(-1)^r T^\mu_1 \exp \left( i\pi \sum_{l=2}^{r-1} T_l^\mu \right) T^\mu_r, 
\end{eqnarray}
with $\mu=x,y,z$. 
One can see that the fully polarized ferromagnetic state, say $\left| 1111 \cdots \right>$ in the $T^x$ basis, has a perfectly saturated string correlation $\langle \tilde{O}^x_\textrm{string} (r) \rangle=-1$. 
Therefore, the corresponding correlation function in the original model also has a perfectly saturated value $\langle (-1)^r \lambda^1_1 \lambda^1_r \rangle =-1$. 
Of course, an exact ground state of the strong-coupling Hamiltonian \eqref{eq:Heff_2nd} is not in the fully polarized state.
However, since the ground state is obtained from the SU(2) ferromagnet perturbed by the $S_4$ anisotropy, it is still very close to the fully polarized state as long as the higher-order perturbations are small. 
A finite expectation value of $\langle (-1)^i \lambda^1_i \lambda^1_i \rangle$ indicates the \emph{staggered} spin imbalance order associated with $(-1)^i \mu^{1234}_i$.

A direct way to see this order is to apply the nonlocal transformation $\calV$ to $\left| 1111 \cdots \right>$. 
$\calV$ acts as $\left| 1 \right> \rightarrow (\left| + \right> + \left| - \right>)/\sqrt{2}$ on odd site, but $\left| 1 \right> \rightarrow (\left| + \right> - \left| - \right>)/\sqrt{2}$ on even site (if the state $\left| 0 \right>$ is inserted, this transformation becomes slightly more complicated). 
This gives the product of the two local states $\left| \Psi_{13} \right>$ and $\left| \Psi_{24} \right>$ on alternating sites. 
These local states are actually the eigenstates of $\mu^{1234}_i$ with eigenvalue $\pm 1$. 
The state $\left| -1 -1 \cdots \right>$ is also transformed to the product of the two states $\left| \Psi_{13} \right>$ and $\left| \Psi_{24} \right>$ but in the opposite manner to $\left| 1111 \cdots \right>$. 
One can repeat similar arguments for the other four states aligned in $y$ and $z$ and obtain the spin imbalanced states corresponding to $(-1)^i \mu^{2314}_i$ and $(-1)^i \mu^{3124}_i$. 
Consequently, we have the almost quantized, or equivalently, almost factorizable spin imbalance state with six-fold degeneracy as shown in Fig.~\ref{fig:SixGS}~(b). 

%%%%%%%%%%%%%%%%%%%%%%%%%%%%%%%%%%%%%%%%%%%%%%%%%%%%%%%%%%%%%%%%%%%%%%%%%%%%%%%%%%%%%%%%%%

\subsection{Ground state of strong-coupling Hamiltonian}

From now on, we confirm the above expectation on the ground state of the strong-coupling Hamiltonian. 
We separately treat the three regimes: (i) $J_d <0$, (ii) $J_d >0$, and (iii) $J_d=0$.

\subsubsection{Regime \texorpdfstring{$J_d<0$}{Jd < 0}}

When $J_d<0$, the ground state is polarized into the single tetramer state $\left| 0 \right>$ on each site. 
Thus we have a unique disordered ground state with a finite excitation gap, where all correlation functions decay exponentially. 
In Ref.~\onlinecite{Cabra1998}, Cabra \textit{et al.} studied the magnetic phase diagram of a four-leg spin tube corresponding to the $J_d=-J_\perp$ case. 
In the weak coupling limit $J_\perp \ll J_\parallel$, they analyzed the model by bosonization and found a possible gapped phase in the 1/4-magnetization plateau. 
That phase is described by the massive sine-Gordon model, whose potential has only a single minimum in the compactification radius, and expected to be unique and disordered. 
Therefore the unique disordered ground state extends from the weak- to strong-coupling regime. 

%%%%%%%%%%%%%%%%%%%%%%%%%%%%%%%%%%%%%%%%%%%%%%%%%%%%%%%%%%%%%%%%%%%%%%%%%%%%%%%%%%%%%%%%%%

\subsubsection{Regime \texorpdfstring{$J_d>0$}{Jd > 0}} \label{sec:SCHam_S12_Jdgreater}

If the diagonal asymmetry is sufficiently strong, $J_d \gg 0$, the two states $\left| + \right>$ and $\left| - \right>$ are energetically favored on each site, 
while exchange processes involving the $\left| 0 \right>$ state will be suppressed. 
In this case, the effective Hamiltonian~\eqref{eq:Heff_asym} takes the following form, 
\begin{eqnarray} \label{eq:Ham_l1l1}
H_\textrm{eff}^{(1)} = \frac{J_\parallel}{4} \sum_{i=1}^L \lambda^1_i \lambda^1_{i+1}. 
\end{eqnarray}
If we regard the $\left| \pm \right>$ states as eigenstates of the pseudo-spin-1/2 operator $\tau^z$ with eigenvalues $\pm 1/2$ 
and neglect the $\left| 0 \right>$ state, this model is nothing but an Ising model in the $x$ direction
\begin{eqnarray} \label{eq:Ham_tauxtaux}
H_\textrm{eff}^{(1)} = J_\parallel\sum_{i=1}^L \tau^x_i \tau^x_{i+1},
\end{eqnarray}
where $\vec{\tau}_i$ is a spin-1/2 operator. 
Thus we obtain a two-fold degenerate ground state, like an Ising N\'eel state, characterized by a finite expectation value of $(-1)^i \lambda^1_i$ (or equivalently, $(-1)^i \tau^x_i$). 
Of course, close to the symmetric point $J_d=0$, the exchange processes involving $\left| 0 \right>$ should be taken into account. 
As discussed in Sec.~\ref{sec:Hidden_ferro}, the field $\lambda^8_i$ acts as an easy-axis anisotropy $-(T^x_i)^2$ on the ferromagnet. 
Thus the ferromagnetic order in the $x$ direction is favored. 
Even in the vicinity of $J_d=0$, this leads the almost quantized expectation value, $\langle (-1)^i \lambda^1_i \rangle = \pm 1$, as if in the classical N\'eel state. 
If we translate the above ground-state properties back in the original tube variables, this indicates a staggered spin imbalance associated with the order parameter $(-1)^i \mu^{1234}_i$. 
This order parameter possesses the symmetry under $C_{2v}=\left\{ (), (13), (24), (13)(24) \right\}$ as a subgroup of the $C_{4v}$. 
Since $C_{4v}/C_{2v} = \mathbb{Z}_2$, this order parameter is compatible with two-fold degeneracy of the ground state. 
The resulting phase is illustrated in Fig.~\ref{fig:SixGS}~(b) in the ``$\lambda^1$ direction.'' 

When increasing $J_{\parallel}$, new terms appear in the Hamiltonian and we obtain, considering only nearest-neighbour tems, an $\mathrm{XYZ}$ model at the second order
\begin{eqnarray} \label{eq:Ham_xyzS1_2}
H_\textrm{eff}^{(2)} = \sum_{i=1}^L \left( J_x \tau^x_i \tau^x_{i+1} + J_y \tau^y_i \tau^y_{i+1} + J_z \tau^z_i \tau^z_{i+1} \right). 
\end{eqnarray}
where $J_y$ and $J_z$ are negative and of order $J_{\parallel}^2/J_{\perp}$ (given by complicated analytical expression). 
Once projected on the truncated subspace, the relations $\tau^y_j=2\chi^{1234}_j$ and $\tau^z_j/2 = Q^{1324}_j$ hold. 
One can check that the form of the Hamiltonian \eqref{eq:Ham_xyzS1_2} is invariant under the $D_2 \times T$ symmetry operations coming from the original $C_{4v}$ symmetry. 
It turns out that, in the regime where the perturbation theory is valid, 
the $J_x$ term always dominates thus we do not expect a transition out of the Ising phase as long as the magnetization plateau exists. 
However, as we will see in Sec.~\ref{sec:GeneralS}, this is not the case for $S>1/2$. 

%%%%%%%%%%%%%%%%%%%%%%%%%%%%%%%%%%%%%%%%%%%%%%%%%%%%%%%%%%%%%%%%%%%%%%%%%%%%%%%%%%%%%%%%%%

\subsubsection{Point \texorpdfstring{$J_d=0$}{Jd = 0}}\label{sec:SCHam_S12_Jd0}

At the $S_4$ symmetric point $J_d=0$, the effective Hamiltonian is given by Eq.~\eqref{eq:Heff_2nd}. 
In this case, we expect the six-fold ground state with the staggered spin imbalance order, as illustrated in Fig.~\ref{fig:SixGS}~(b), associated with the three order parameters $(-1)^i\lambda^1_i$, $(-1)^i\lambda^4_i$, and $(-1)^i\lambda^6_i$. 
These operators transform each other by the $\mathbb{Z}_3$ symmetry operation in Eq.~\eqref{eq:C3op}. 
Although the dimer states $\left| \Psi_{jk} \right>$ are not orthogonal between each other ($|\Psi_{13}\rangle$ is only orthogonal to $|\Psi_{24}\rangle$ for example), 
the overlaps of the six product states built from them scale as $1/2^L$, similar to valence-bond solid states.~\cite{Affleck1987,Oshikawa1992} 
Therefore, the six ground states are not orthogonal in a finite system but asymptotically orthogonal in the thermodynamic limit $L \rightarrow \infty$. 

\begin{figure}
\includegraphics[width=0.45\textwidth,clip]{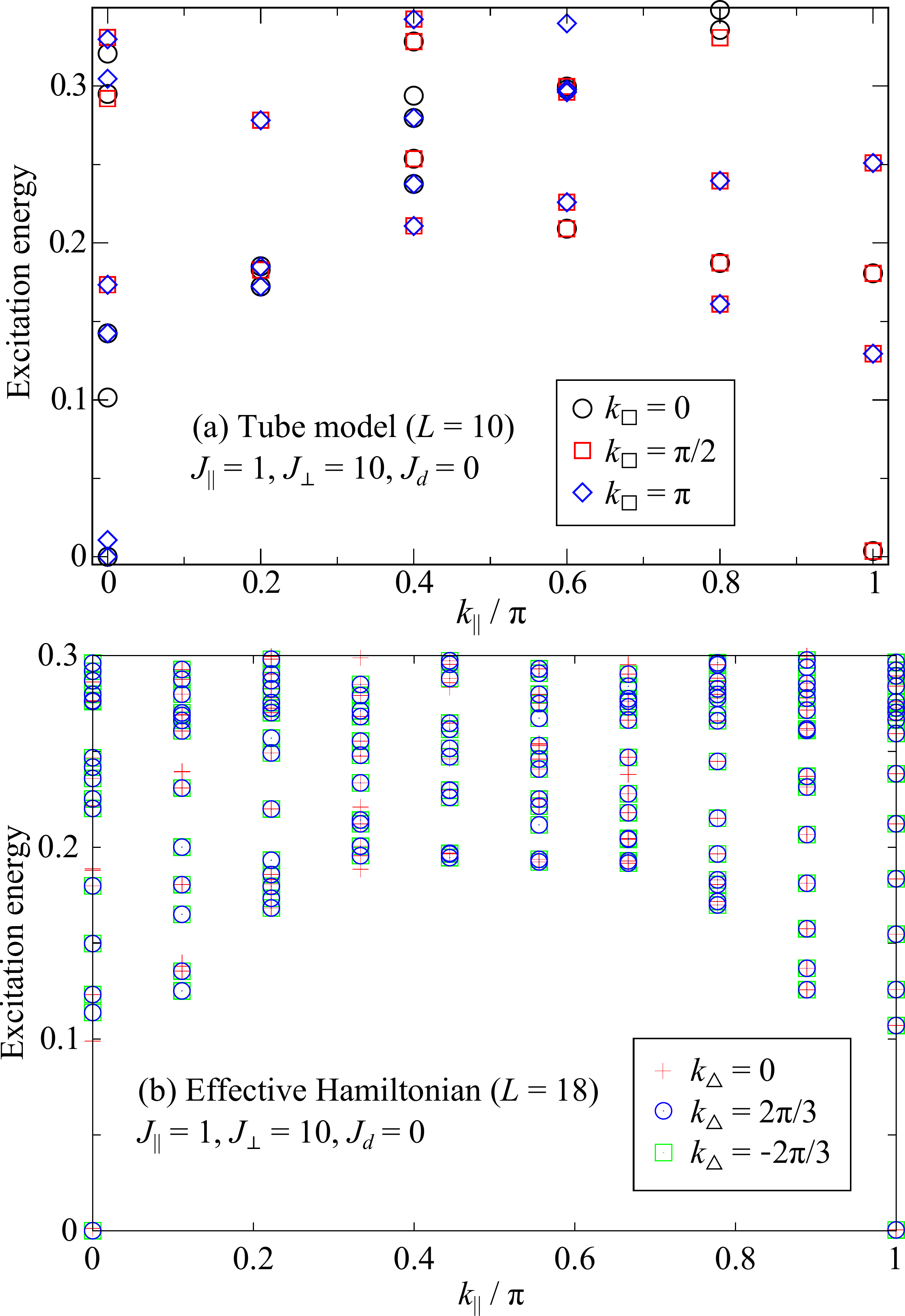}
\caption{(Color online) Excitation energies are plotted as functions of longitudinal momentum $k_{\parallel}$ for $J_\parallel=1$, $J_\perp=10$, and $J_d=0$. 
The top panel (a) shows ED data obtained from the tube model with $L=10$ and labeled by the transverse momenta $k_\square$. 
The bottom panel (b) shows the ED data obtained from the second-order effective Hamiltonian~\eqref{eq:Heff_2nd} with $L=18$ and labeled by $k_\triangle$.}
\label{fig:dispersion}
\end{figure}

Since there is no analytical way to handle the effective Hamiltonian~\eqref{eq:Heff_2nd}, 
we first examine it numerically in order to support the above proposal. 
Using exact diagonalization (ED) technique, we compute the low-lying excitation energies for the original tube model~\eqref{eq:4leg_tube_Ham} with $L=10$ 
and the effective Hamiltonian with $L=18$, at $J_\parallel=1$, $J_\perp=10$, and $J_d=0$. 
They are shown in Fig.~\ref{fig:dispersion} as functions of the longitudinal momentum $k_{\parallel}$ since we impose the PBC in the leg direction. 
Both results are in good quantitative agreement and exhibit a six-fold (nearly) degenerate structure in the lowest energies and a large gap above them. 
This is a strong evidence of the ground state with discrete $\mathbb{Z}_6$ symmetry breaking. 

In the ED calculation on the second-order effective Hamiltonian~\eqref{eq:Heff_2nd}, we implemented the global $\mathbb{Z}_3$ symmetry as well as the longitudinal translational symmetry. 
The excitation spectrum is resolved by $k_\triangle$ which is defined by $\prod_i \mathcal{X}_i \left| k_\triangle \right> = \exp(ik_\triangle) \left| k_\triangle \right>$ 
and take three values, $0$ and $\pm 2\pi /3$. 
As seen from Fig.~\ref{fig:dispersion} (b), the six ground states belong to each six symmetry sector characterized by $k_\triangle$ and $k_{\parallel}=0,\pi$. 
This observation is consistent with one-dimensional irreducible representations of the $\mathbb{Z}_3$ and translational symmetries, 
formed by linear combinations of the six spin imbalance states displayed in Fig.~\ref{fig:SixGS} (b). 
On the other hand, for the diagonalization on the original Hamiltonian, we implement the global $\mathbb{Z}_4$ symmetry associated with the cyclic permutation of legs 
and classify the spectrum by the momentum $k_\square$. We can also access the reflection quantum numbers ${\bf R}=(r_x,r_y)$ 
labeling the even/odd states with respect to reflections respectively along the leg and rung directions.~\footnote{In principle, for $J_d=0$, the symmetry 
group is $S_4$ but we have only considered the subgroup $C_{4v}=\mathbb{Z}_4 \times \mathbb{Z}_2$ which is (i) the point-group symmetry for $J_d \neq 0$; (ii) 
easier to implement with one-dimensional irreducible representations only}
If we denote each symmetry sector as ${\bf K}=(k_{\parallel},k_\square)$, from Fig.~\ref{fig:dispersion} (a), 
we can find that the six lowest-energy states have quantum numbers:
\begin{itemize}
\item[$\bullet$] ${\bf K}=(0,0)$ and ${\bf R}=(+,+)$ (2 states)
\item[$\bullet$] ${\bf K}=(0,\pi)$ and ${\bf R}=(-,+)$ (1 state)
\item[$\bullet$] ${\bf K}=(\pi,\pi)$ and ${\bf R}=(+,+)$ (1 state)
\item[$\bullet$] ${\bf K}=(\pi,\pi/2)$ and ${\bf R}=(N.A.,+)$ (1 state)
\item[$\bullet$] ${\bf K}=(\pi,-\pi/2)$ and ${\bf R}=(N.A.,+)$ (1 state)
\end{itemize}
where $N.A.$ stands for not available (symmetries not commuting). This is again compatible with the irreducible representations of the $C_{4v}=\mathbb{Z}_4 \times \mathbb{Z}_2$ and translational symmetries. 

\begin{figure}
\includegraphics[width=0.45\textwidth,clip]{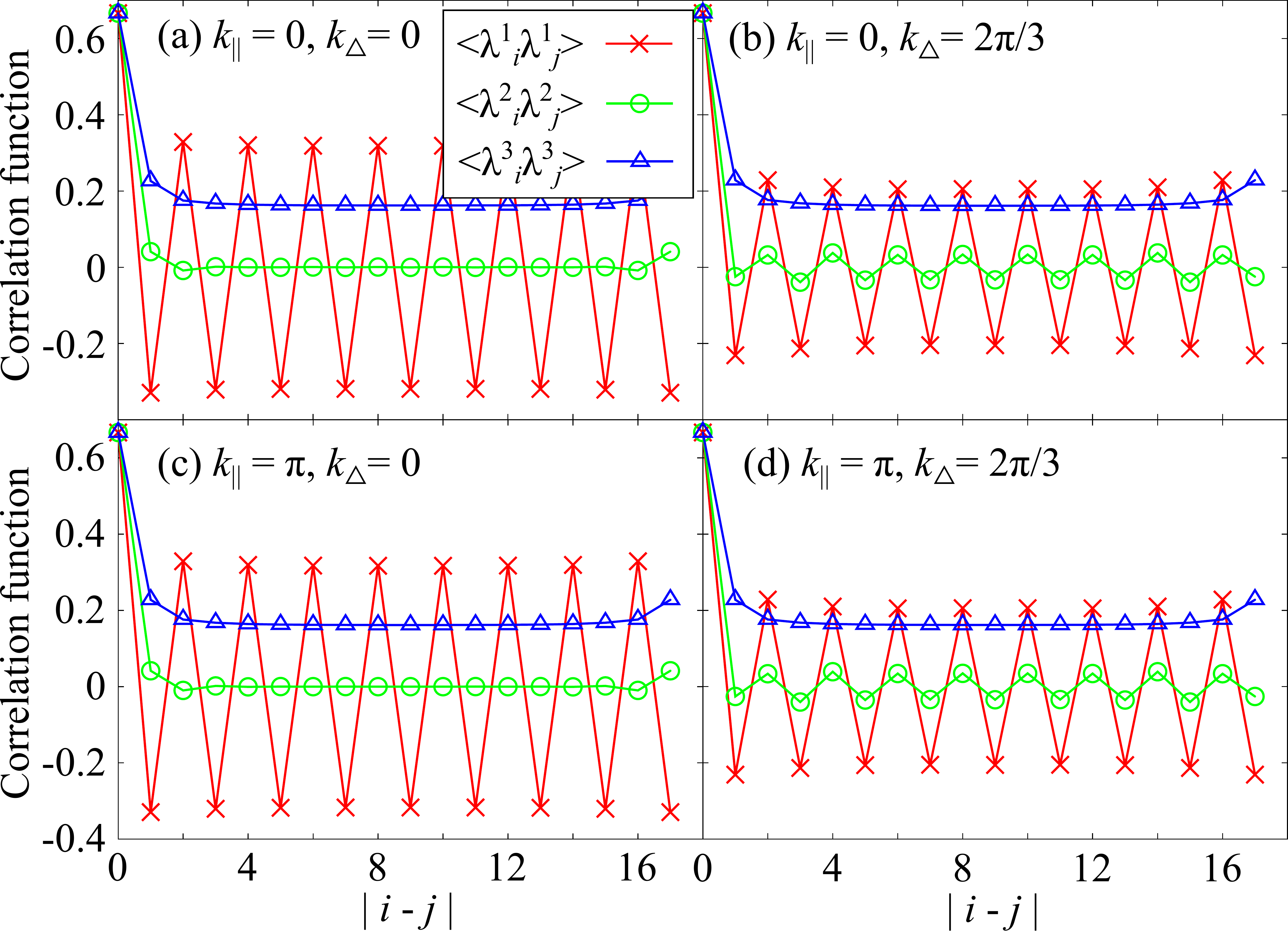}
\caption{(Color online) Correlation functions with respect to the ground state of the the second-order effective Hamiltonian~\eqref{eq:Heff_2nd} for $L=18$ and $J_\parallel/J_\perp=0.1$. 
The results are obtained for each symmetry sector (a) $k_{\parallel}=0$, $k_\triangle=0$, (b) $k_{\parallel}=0$, $k_\triangle=2\pi/3$, (c) $k_{\parallel}=\pi$, $k_\triangle=0$, and (d) $k_{\parallel}=\pi$, $k_\triangle=2\pi/3$. 
The cross, circle, and triangle symbols denote $\left< \lambda^1_i \lambda^1_j \right>$, $\left< \lambda^2_i \lambda^2_j \right>$, 
and $\left< \lambda^3_i \lambda^3_j \right>$, respectively.}
\label{fig:CF_ED_Heff}
\end{figure}

We also calculate the correlation functions, $\langle \lambda^1_i \lambda^1_j \rangle$, $\langle \lambda^2_i \lambda^2_j \rangle$, and $\langle \lambda^3_i \lambda^3_j \rangle$, 
with respect to the ground state in each symmetry sector $(k_{\parallel},k_\triangle)$ for the effective Hamiltonian~\eqref{eq:Heff_2nd}, which are shown in Fig.~\ref{fig:CF_ED_Heff}. 
Oscillating behaviors in $\langle \lambda^1_i \lambda^1_j \rangle$ indicates the staggered spin imbalance 
while strong suppressions of $\langle \lambda^2_i \lambda^2_j \rangle$ means no development of the spin vector chiral order. 
Although $\langle \lambda^3_i \lambda^3_j \rangle$ exhibits a finite uniform correlation, this does not necessarily indicate the existence 
of another order associated with $\lambda^3$. Since we can write 
\begin{eqnarray} \label{eq:L3_L6L4}
\lambda^3 = \frac{1}{2} \left[ \exp (i\pi \lambda^6) -\exp (i\pi \lambda^4) \right], 
\end{eqnarray}
$\lambda^3_i$ becomes $+1/2$ $(-1/2)$ if $\lambda^4_i$ takes $\pm 1$ $(0)$ and $\lambda^6_i$ takes $0$ $(\pm 1)$ as in the spin imbalance phase. 
Combined with the fact that the degenerate ground state obtained by ED is in a superposition of the six spin imbalance states to respect the $\mathbb{Z}_3$ and translational symmetries, 
this gives the finite values of $\langle \lambda^3_i \lambda^3_j \rangle$ in addition to $\langle \lambda^1_i \lambda^1_j \rangle$. 
Overall, for the $m=1/4$ plateau, our ED data strongly suggest the realization of the six-fold degenerate ground state with staggered spin imbalance 
in the strong-coupling limit. 
However the quantization of the order parameter cannot be observed due to the limitation of the system size, impotant because of the non-orthogonality of the degenerate spin imbalance states. 
We will confirm this picture and further address the quantized spin imbalance order in Sec.~\ref{sec:num} on the original tube systems with large-scale simulations.

%%%%%%%%%%%%%%%%%%%%%%%%%%%%%%%%%%%%%%%%%%%%%%%%%%%%%%%%%%%%%%%%%%%%%%%%%%%%%%%%%%%%%%%%%%
%%%%%%%%%%%%%%%%%%%%%%%%%%%%%%%%%%%%%%%%%%%%%%%%%%%%%%%%%%%%%%%%%%%%%%%%%%%%%%%%%%%%%%%%%%

\subsection{General \texorpdfstring{$S$}{S}: highest plateau} \label{sec:GeneralS}

Finally, we consider the higher $S$ cases. For a generic magnetization plateau, the strong-coupling Hamiltonian approach becomes
too difficult to handle because of the increasing number of low-energy states. Yet, our preceding discussions on the $S=1/2$ case can be directly applied to 
these cases in the highest plateau (not counting the saturated plateau) of magnetization per spin $m=S-1/4$. 
When solving the single tetrahedron, there are four eigenstates which can be written exactly 
as Eq.~\eqref{eq:eigenstates_chi} with the changes $\uparrow \to S$, $\downarrow \to S-1$. We now show that, for any $S > 1/2$, this leads to the appearance of 
new phases, for both cases $J_d = 0$ and $J_d > 0$.

%%%%%%%%%%%%%%%%%%%%%%%%%%%%%%%%%%%%%%%%%%%%%%%%%%%%%%%%%%%%%%%%%%%%%%%%%%%%%%%%%%%%%%%%%%

\subsubsection{Regime \texorpdfstring{$J_d>0$}{Jd > 0}} \label{sec:SCHam_genS_Jdgreater}

\begin{figure}
\includegraphics[width=0.46\textwidth,clip]{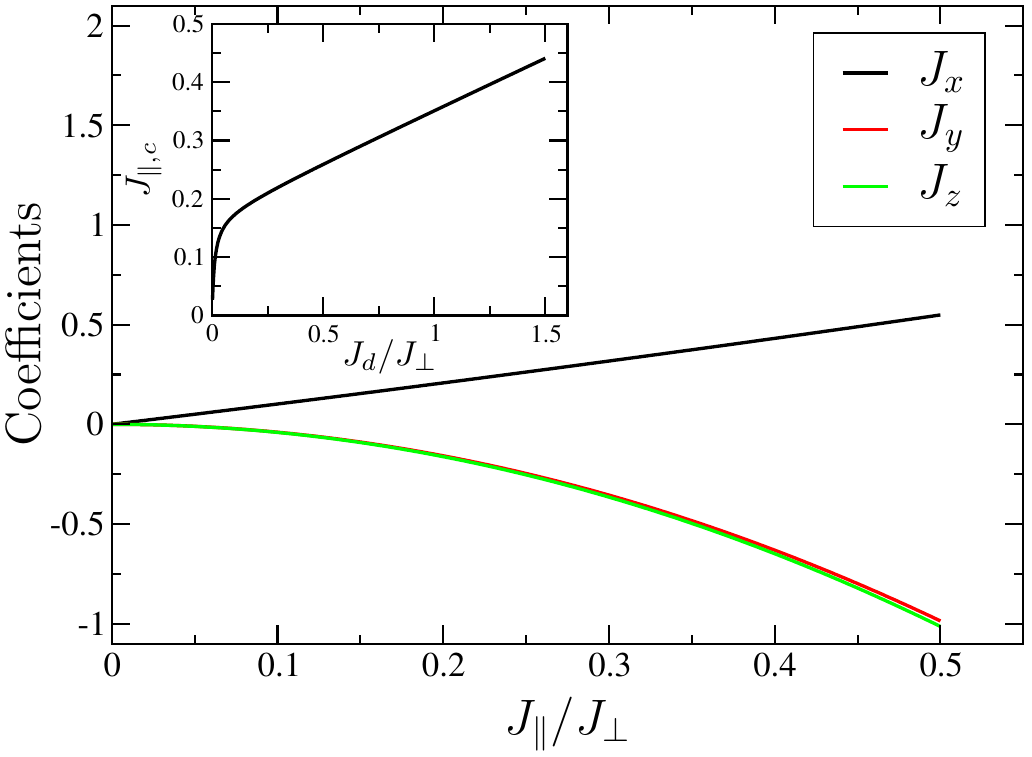}
\caption{(Color online) Values of the parameters $J_x$, $J_y$, and $J_z$ of the effective XYZ second-order Hamiltonian (\ref{eq:Ham_xyzS1_2}) as functions of $J_\parallel/J_\perp$ for $S=1$ and $J_d/J_\perp=0.5$. The inset shows the value of the critical coupling $J_{\parallel,c}$, defined by $J_z/J_x=-1$, as a function of $J_d/J_{\perp}$.}
\label{fig:ParametersXYZS1Jd05}
\end{figure}

The second-order effective Hamiltonian in this general spin-$S$ case is an $\mathrm{XYZ}$ model as in Eq.~\eqref{eq:Ham_xyzS1_2} with couplings being complicated functions of
$J_{\perp}$, $J_d$, and $S$. We plot in Fig.~\ref{fig:ParametersXYZS1Jd05} the values of those couplings as functions of $J_{\parallel}$ for $S=1$ and $J_d=0.5$. Several comments
have to be made.

First, contrary to the $S=1/2$ case where $J_x$ is always dominant coupling, $|J_z| \sim |J_y| > |J_x|$ occurs even in the perturbative regime $J_{\parallel}/J_{\perp}\lesssim 1$. 
Then, we expect a transition from an antiferromagnetic Ising phase where the positive coefficient $J_x$ dominates to a ferromagnetic Ising phase where one of the negative $J_y$ or $J_z$ has the largest magnitude. 
From Fig.~\ref{fig:ParametersXYZS1Jd05} for $S=1$, $J_z$ dominates (it is also true for a higher $S$) but it is difficult to rule out the possibility of having an other regime where $J_y$ becomes smaller, since they take very close values at the second order. 
Higher-order terms possibly lead to an extra transition appearing if their values cross for larger $J_{\parallel}$. 
However, from the results at $J_d=0$ (see below), it appears that the $J_z$ coupling always dominates. 
We observe a uniform ordering of the $Q^{1324}$ operator where all the sites are either in the tetramer state $|+\rangle$ or in $|-\rangle$ (see Eq.~\eqref{eq:eigenstates_X} and Fig.~\ref{fig:Ising_phase_cartoon}~(b)). 
On the other hand, we do not find any sign of chiral order. 
Like the $\mu_i^{1234}$ order parameter, the $Q^{1324}$ operator now possesses the order-4 symmetry, but in a different way, namely $\left\{ (), (12)(34), (14)(23), (13)(24) \right\}$. 
This leads the two-fold degenerate ground state with the uniform tetramer order.

Since our effective Hamiltonian \eqref{eq:Ham_xyzS1_2} is of the form of an XYZ model, the transition passes through the U(1) symmetric point $J_x=-J_z$. 
Apparently the transition becomes the continuous one with the central charge $c=1$. 
Of course, this is merely due to the truncation of higher-order perturbations; including those perturbations, this emergent U(1) symmetry will be broken. 
In general, between two ordered phases associated with different order parameters, there is a first-order transition or an intermediate phase where both order parameters coexist. 
However, several exceptions of this criterion exist in 1D due to strong quantum fluctuation. 
Indeed, even in the absence of an exact U(1) symmetry, we still have a Gaussian transition with $c=1$ under the dihedral group symmetry of two spin axes, provided by the $C_{4v}$ symmetry. 
This is our case and the transition becomes continuous although both phases have different symmetries. 

%%%%%%%%%%%%%%%%%%%%%%%%%%%%%%%%%%%%%%%%%%%%%%%%%%%%%%%%%%%%%%%%%%%%%%%%%%%%%%%%%%%%%%%%%%

\subsubsection{Point \texorpdfstring{$J_d=0$}{Jd = 0}}

Moving to the symmetric point, the second-order effective Hamiltonian in the strong-coupling limit is given by the same form as 
Eq.~\eqref{eq:Heff_2nd}, except for the $S$-dependent coupling constants, 
\begin{eqnarray} \label{eq:Coupling_GenS}
&& q_1 = \frac{J_\parallel}{4}-\frac{J_\parallel^2}{J_\perp} \left( \frac{32S^2-16S-3}{32} +\frac{5}{256S} \right), \nonumber \\
&& q_2 = -\frac{J_\parallel^2}{J_\perp} \left( \frac{32S^2-1}{32} +\frac{3}{256S} \right), \nonumber \\
&& q_3 = -\frac{J_\parallel^2}{J_\perp} \left( \frac{32S^2+1}{32} -\frac{3}{256S} \right), \nonumber \\
&& t_1 = -\frac{J_\parallel^2}{32J_\perp}, \hspace{10pt}
t_2 = -\frac{J_\parallel^2}{48J_\perp}. 
\end{eqnarray}

At the first order in $J_\parallel$, we obtain exactly the same Hamiltonian as Eq.~\eqref{eq:Heff_1st} 
and therefore the hidden SU(2) symmetry causes the macroscopic degeneracy in the ground state. 
Again, adding the second-order perturbations, we will find the staggered spin imbalance phase 
associated with a finite expectation value of $(-1)^i\lambda^{1,4,6}_i$. 
In the $S=1/2$ case, this follows from the fact that $q_1$ is positive and always larger than other couplings 
in its magnitude for the strong-coupling regime $J_\parallel/J_\perp \ll 1$. 

\begin{figure*}
\includegraphics[width=0.9\textwidth,clip]{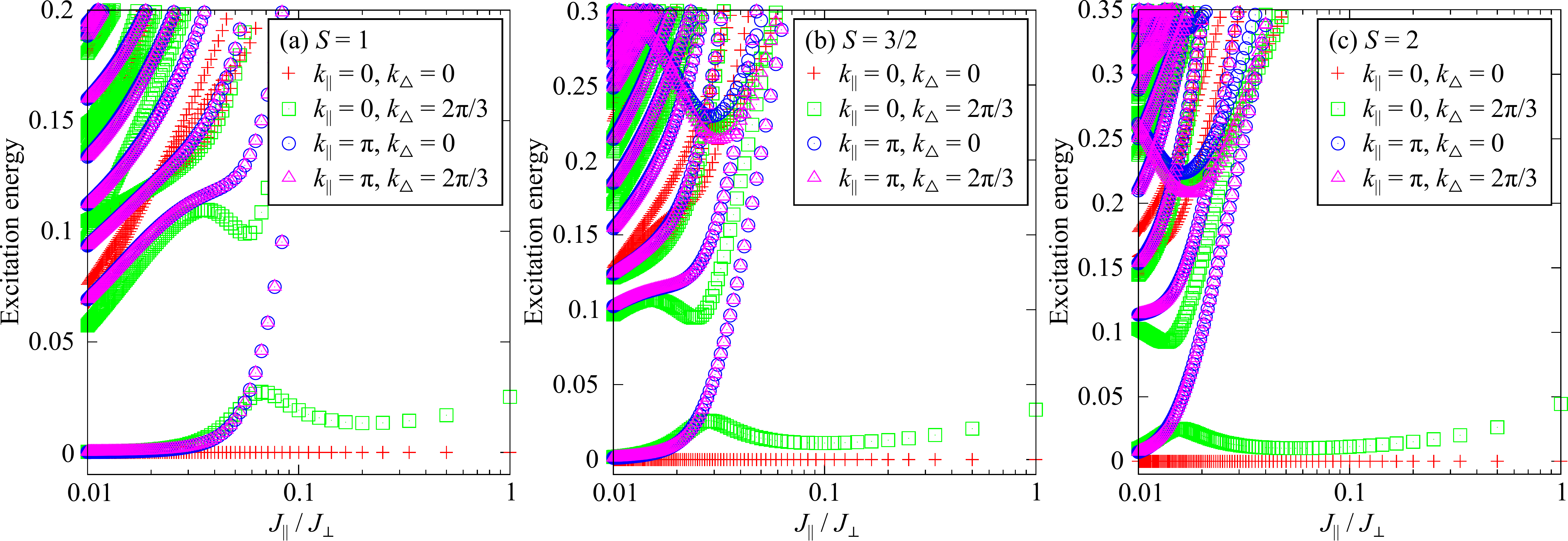}
\caption{(Color online) Excitation energies obtained in the effective Hamiltonian~\eqref{eq:Heff_2nd} 
are plotted against $J_\parallel/J_\perp$ for (a) $S=1$, (b) $S=3/2$, and (c) $S=2$. 
The logarithmic scale is used for the horizontal axis. 
Here we set $J_\parallel=1$ and use the $L=16$ system. 
Each symbol corresponds to the lowest energy eigenvalues associated with the set of quantum numbers: 
$k_{\parallel}=0$ and $k_\triangle=0$ (cross), $k_{\parallel}=0$ and $k_\triangle=2\pi/3$ (square), $k_{\parallel}=\pi$ and $k_\triangle=0$ (circle), 
and $k_{\parallel}=\pi$ and $k_\triangle=2\pi/3$ (triangle). 
Energy levels of the $k_\triangle=2\pi/3$ sector are degenerate with those of $k_\triangle=-2\pi/3$.}
\label{fig:ED_Heff_GenS}
\end{figure*}

However, this is no longer true for $S>1/2$ cases. 
Increasing $J_\parallel/J_\perp$ from zero, we can find a regime where $q_3$ becomes the negative most dominant coupling. 
This implies that another ordered phase associated with $\lambda^3_i$ or $\lambda^8_i$ is possible to occur along $J_\parallel/J_\perp$. 
In Fig.~\ref{fig:ED_Heff_GenS}, we show the lowest excitation energies for the effective Hamiltonian~\eqref{eq:Heff_2nd} with the coupling constants~\eqref{eq:Coupling_GenS} for several $S$. 
Since we are interested in the ground state, it is enough to look at the excitation spectra at $k_{\parallel}=0$ and $k_{\parallel}=\pi$. 
One should notice that each spectrum with $k_\triangle=2\pi/3$ is doubly degenerate with that with $k_\triangle=-2\pi/3$. 
As expected from the $S=1/2$ case, the (nearly) six-fold degenerate energy corresponding to the staggered spin imbalanced phase lies around $J_\parallel/J_\perp=0.01$. 
Increasing $J_\parallel/J_\perp$, the three lowest energies with $k_{\parallel}=\pi$ are lifted while the other three with $k_{\parallel}=0$ still remain. 
This indicates that a uniform ordered phase with $\mathbb{Z}_3$ symmetry breaking appears in the intermediate coupling regime.

In fact, this corresponds to a three-fold degenerate ground state with uniform tetramer order associated with $\lambda^8_i$ and its $\mathbb{Z}_3$ symmetry counterparts $\mathcal{X} \lambda^8_i \mathcal{X}^{-1}$ and $\mathcal{X}^2 \lambda^8_i \mathcal{X}^{-2}$. 
In the original tube, these order parameters correspond to the plaquette operators $P^{1234}_i$, $P^{3124}_i$, and $P^{2314}_i$ defined in Table~\ref{table:OrderParam}, respectively. 
Since $P^{jklm}_i$ preserves the order-8 symmetry, this clearly detects the $\mathbb{Z}_3$ symmetry breaking. 
In the above $J_d>0$ case, since the $S_4$ symmetry is initially broken, $Q^{1324}_i$ is equivalent to $P^{3124}_i$ or $P^{2314}_i$ in the sense of order parameter which detects $\mathbb{Z}_2$ symmetry breaking. 
Nevertheless, for $J_d=0$, we can still use \emph{two} independent $Q^{jklm}_i$ instead of $P^{jklm}_i$ to detect the tetramer order as indicated in Eq.~\eqref{eq:PandQ}. 
Namely, the same magnitude of expectation values of two different $Q^{jklm}_i$ implies an additional $\mathbb{Z}_2$ symmetry and detect the tetramer order. 
For the simplest three product states with maximal tetramer order, $\left| \Psi_\nu \right> = \bigotimes_i \left| \nu \right>_i$, $\nu=+,-,0$ in the Q basis, $Q^{jklm}_i$ takes the expectation values indicated in Table~\ref{tab:Q_averages}. 

\begin{table}
    \begin{ruledtabular}
    \begin{tabular}{l|ccc}
         & $Q_i^{1234}$ & $Q_i^{1324}$ & $Q_i^{1423}$ \\
        \hline
        $| \Psi_{+} \rangle$  & 0 & $2S^3$ & $2S^3$ \\
        $| \Psi_{-} \rangle$  & $-2S^3$ & $-2S^3$ & 0 \\
        $| \Psi_{0} \rangle$  & $2S^3$ & 0 & $-2S^3$ \\
    \end{tabular}
    \end{ruledtabular}
    \caption{Expectation values of the $Q_i^{jklm}$ operators in the tetramer ordered phase.}
    \label{tab:Q_averages}
\end{table}

Although the two phases appearing in this strong-coupling regime are understood, the question of the transition is actually complicated. 
Again the two phases have different order parameters. 
The standard Landau theory generally tells us that there is a first-order transition or an intermediate phase with coexistence of the order parameters. 
However, as seen in the $J_d>0$ case, we cannot exclude the possibility of a continuous transition. 
We could not extract any information about the nature of the transition from the second-order Hamiltonian \eqref{eq:Heff_2nd}. 
In the next section, we will provide numerical results supporting a continuous scenario.

%%%%%%%%%%%%%%%%%%%%%%%%%%%%%%%%%%%%%%%%%%%%%%%%%%%%%%%%%%%%%%%%%%%%%%%%%%%%%%%%%%%%%%%%%%
%%%%%%%%%%%%%%%%%%%%%%%%%%%%%%%%%%%%%%%%%%%%%%%%%%%%%%%%%%%%%%%%%%%%%%%%%%%%%%%%%%%%%%%%%%
%%%%%%%%%%%%%%%%%%%%%%%%%%%%%%%%%%%%%%%%%%%%%%%%%%%%%%%%%%%%%%%%%%%%%%%%%%%%%%%%%%%%%%%%%%

\section{DMRG results}\label{sec:num}

We here use the standard DMRG algorithm~\cite{White1992} to investigate physical properties on the magnetization plateau in the original spin tube \eqref{eq:4leg_tube_Ham}. Typically, when computing energies or local quantities, we have kept 1600 states (respectively 3200 states) for $S=1/2, 1$ (respectively $S=3/2$) which is sufficient to have a negligible discarded weight (below $10^{-9}$). When computing correlations or entanglement entropies at transitions, it was necessary to keep up to 4000 states to reach convergence. In the following we will set $J_{\perp}=1$.

\subsection{\texorpdfstring{$S=1/2$}{S=1/2}}

\begin{figure}
\begin{center}
\includegraphics[width=\columnwidth,clip]{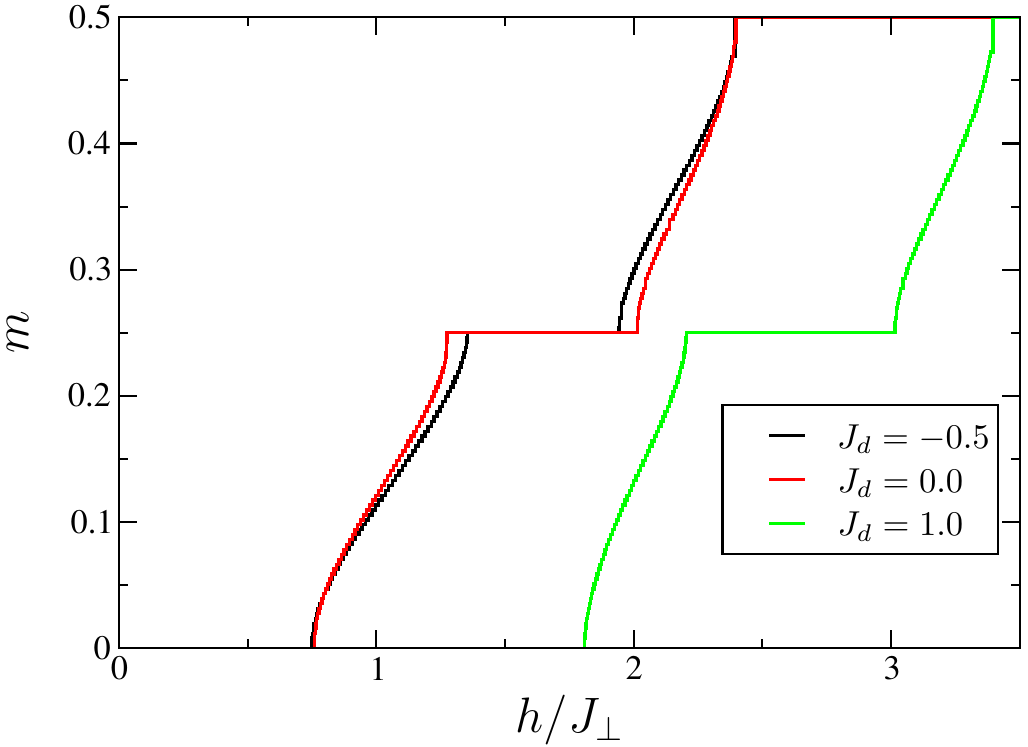}
\end{center}
\caption{(Color online) Magnetization curves obtained by DMRG for several values of $J_d/J_\perp$ in Eq.~\eqref{eq:4leg_tube_Ham} with $S=1/2$, $J_\parallel/J_\perp=0.2$, and $L=64$.}
\label{fig:MofH_S12}
\end{figure}

First of all, by measuring the energy against total $S_z$ and performing a Legendre transformation, we can draw the magnetization curve as plotted in Fig.~\ref{fig:MofH_S12} for $S=1/2$ and $J_\parallel/J_\perp=0.2$. 
Clearly, three magnetization plateaus appear at $m=0$, $m=1/4$, and $m=1/2$; 
The $m=0$ plateau implies a finite triplet excitation gap and the $m=1/2$ plateau corresponds to the fully saturated state. 
Now we are interested in the $m=1/4$ plateau. 
The saturation field $h_{\mathrm{sat}}$ and the spin gap for $J_d<0$ are easily shown to be independent of $J_d$ (for any $S$). 
A finite-size scaling analysis of the $m=1/4$ plateau width does confirm that it remains finite in the thermodynamic limit for all parameters that we study below (data are not shown). 

%%%%%%%%%%%%%%%%%%%%%%%%%%%%%%%%%%%%%%%%%%%%%%%%%%%%%%%%%%%%%%%%%%%%%%%%%%%%%%%%%%%%%%%%%%

\subsubsection{Regime \texorpdfstring{$J_d<0$}{Jd < 0}}

We have verified that for $J_d<0$ we have a unique disordered ground state, by computing both the local quantities and the correlations of the operators defined in Table~\ref{table:OrderParam}. 
Both data are compatible with a unique nondegenerate state, very close to the product of the $\left| 0 \right>$ state, $\left| \Psi_0 \right>$, as expected from the strong-coupling analysis. 
In particular, all connected correlations decay exponentially and for the local magnetizations no spin imbalance is observed (data are not shown). 
This case encompasses the non-frustrated four-leg tube with $J_d=-J_{\perp}$ \cite{Cabra1998}. 

%%%%%%%%%%%%%%%%%%%%%%%%%%%%%%%%%%%%%%%%%%%%%%%%%%%%%%%%%%%%%%%%%%%%%%%%%%%%%%%%%%%%%%%%%%

\subsubsection{Regime \texorpdfstring{$J_d>0$}{Jd > 0}}

\begin{figure}
\begin{center}
\includegraphics[width=\columnwidth,clip]{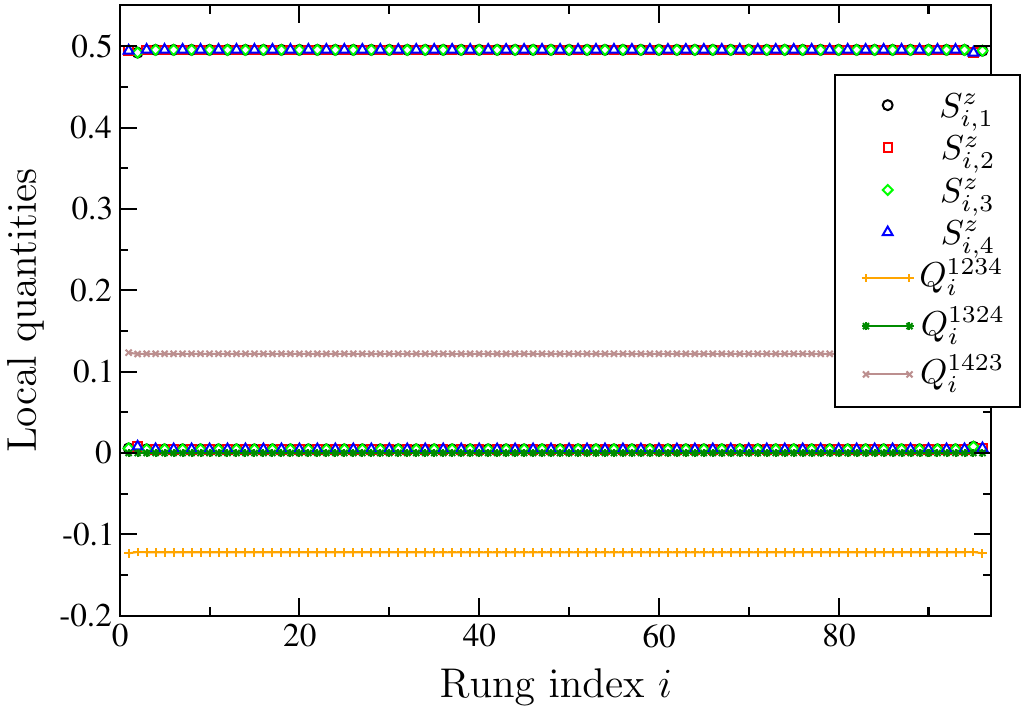}
\end{center}
\caption{(Color online) Local quantities $S^z_{i,j}$ and $Q^{jklm}_i$ as functions of rung index $i$ for $J_d/J_\perp=1$, $J_\parallel/J_\perp=0.1$, and $L=96$. 
The staggered values of $\langle S^z_{i,j} \rangle$ indicate that the simulation selects one of the two spin imbalance states predicted by the strong-coupling analysis.}
\label{fig:Jdsup0_loc_S12}
\end{figure}

Let us now move to the opposite side, namely $J_d>0$. 
In Fig.~\ref{fig:Jdsup0_loc_S12}, we plot the expectation values of the local operators $S^z_{i,j}$ and $Q^{jklm}_i$ for $J_d/J_\perp=1$ and $J_\parallel/J_\perp=0.1$. 
It is obvious that the simulation selects one of the two degenerate ground states \cite{Jiang2013} with the staggered spin imbalance predicted by the strong-coupling analysis from the values of  $\langle S^z_i \rangle$. 
Because of this selection, we can use the local quantities rather than the correlation functions to characterize the ground state.
$\langle S^z_{i,1} \rangle$ and $\langle S^z_{i,3} \rangle$ take the value very close to $+1/2$ on odd plaquettes while $0$ on even ones, and vice versa for $\langle S^z_{i,2} \rangle$ and $\langle S^z_{i,4} \rangle$. 
Then we have the staggered spin imbalance without fluctuation, $\langle (-1)^i \mu^{1234}_i \rangle \simeq -1$. 
This indicates that the ground state is close to a product states of $\left| \Psi_{24} \right>$ on odd plaquettes and $\left| \Psi_{13} \right>$ on even ones. 
The finite expectation value of $Q^{1234}_i = -Q^{1423}_i$ just accompanies the staggered spin imbalance and is very close to $-1/8$ as expected from Eq.~\eqref{eq:L3_L6L4} (recall $\mathcal{P}_i Q^{1324}_i \mathcal{P}_i = \lambda^3_i/4$). 

%%%%%%%%%%%%%%%%%%%%%%%%%%%%%%%%%%%%%%%%%%%%%%%%%%%%%%%%%%%%%%%%%%%%%%%%%%%%%%%%%%%%%%%%%%

\subsubsection{Point \texorpdfstring{$J_d=0$}{Jd = 0}}

\begin{figure}
\begin{center}
\includegraphics[width=\columnwidth,clip]{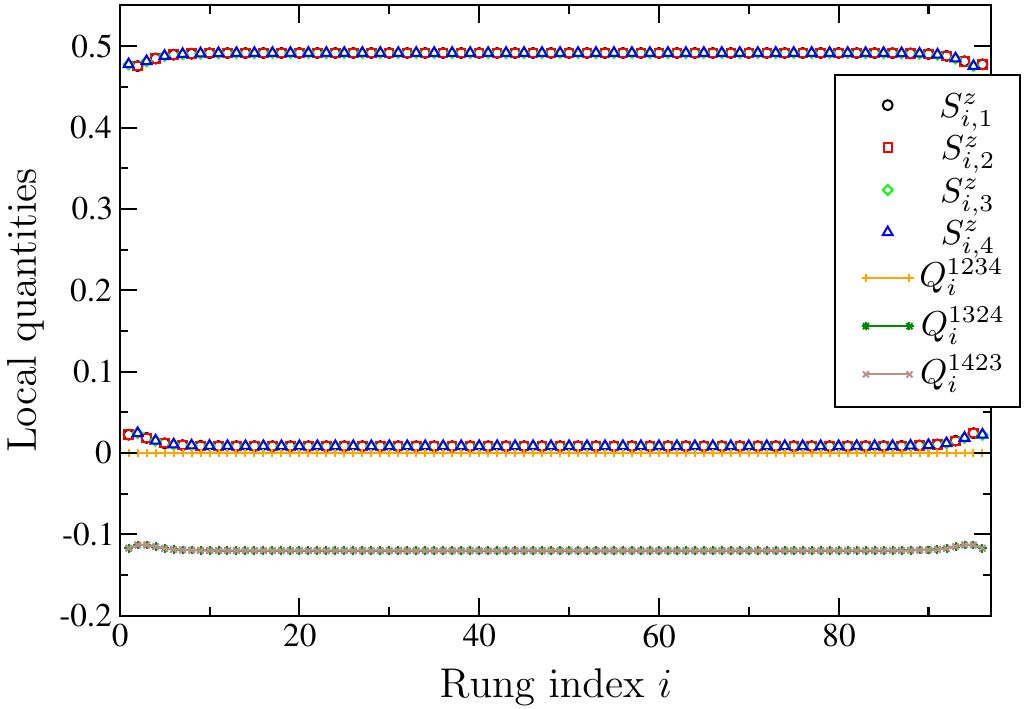}
\end{center}
\caption{(Color online) Local quantities $S^z_{i,j}$ and $Q^{jklm}_i$ as functions of rung index $i$ for $J_d=0$, $J_\parallel/J_\perp=0.1$, and $L=96$.}
\label{fig:Jdeq0_loc_S12}
\end{figure}

Now we are at the $S_4$ symmetric point $J_d=0$. 
We plot in Fig.~\ref{fig:Jdeq0_loc_S12} the local quantities computed for $J_{\parallel}/J_{\perp}=0.1$.
Like for $J_d>0$, the simulation selects one of the six ground states with the staggered spin imbalance pattern. 
Depending on the parameters of the simulation (such as size, or labelling of the 1D path that we use for the simulation), the selected state is not always the same and we have observed several of the six states. 
In Fig.~\ref{fig:Jdeq0_loc_S12}, we observed that $\langle S^z_{i,3} \rangle$ and $\langle S^z_{i,4} \rangle$ take the value very close to $+1/2$ on odd plaquettes while $0$ on even ones, and vice versa for $\langle S^z_{i,1} \rangle$ and $\langle S^z_{i,2} \rangle$. 
This means that $\langle (-1)^i \mu^{2314}_i \rangle \simeq +1$ and the ground state is close to a product state of $\left| \Psi_{12} \right>$ on odd plaquettes and $\left| \Psi_{34} \right>$ on even ones. 
Accompanying the spin imbalance order, $\langle Q^{1324}_i \rangle = \langle Q^{1423}_i \rangle$ takes the value very close to $-1/8$. 

%%%%%%%%%%%%%%%%%%%%%%%%%%%%%%%%%%%%%%%%%%%%%%%%%%%%%%%%%%%%%%%%%%%%%%%%%%%%%%%%%%%%%%%%%%
%%%%%%%%%%%%%%%%%%%%%%%%%%%%%%%%%%%%%%%%%%%%%%%%%%%%%%%%%%%%%%%%%%%%%%%%%%%%%%%%%%%%%%%%%%

\subsection{General \texorpdfstring{$S$ case}{General S case}}

\begin{figure}
\begin{center}
\includegraphics[width=\columnwidth,clip]{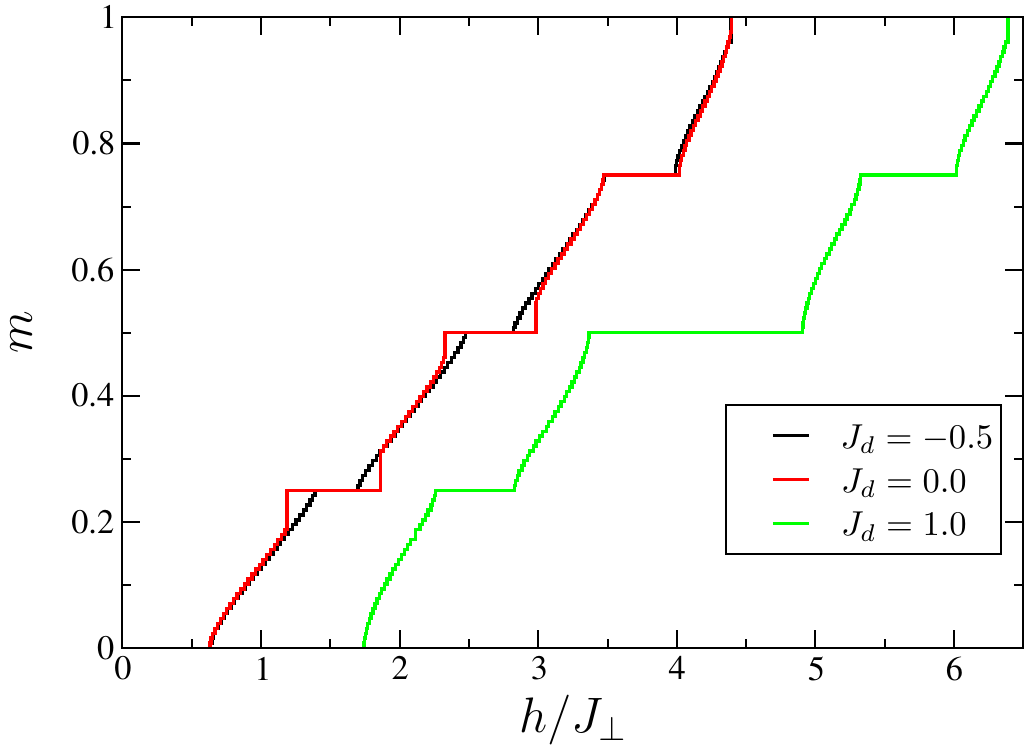}
\end{center}
\caption{(Color online) Magnetization curves obtained by DMRG for several values of $J_d/J_\perp$ in Eq.~\eqref{eq:4leg_tube_Ham} with $S=1$, $J_\parallel/J_\perp=0.1$, and $L=32$.}
\label{fig:MofH_S1}
\end{figure}

We treat now the case of higher spin-$S$ cases: $S=1$ and $3/2$. 
We give the results for the highest plateau to confirm the appearance of another phase with tetramer order in the regime of larger $J_{\parallel}/J_{\perp}$. 
We mainly present results obtained for $S=1$ on the highest plateau, which allows to access larger system sizes in DMRG. 
For completeness, we present in Fig.~\ref{fig:MofH_S1} magnetization curves for $S=1$ and $J_\parallel/J_\perp=0.1$ on a $L=32$ lattice, where the plateaus at $m=1/4$, $1/2$, and $3/4$ clearly appear. 
We note the presence of jumps at the edges of the magnetization plateaus for $m=1/4$ and $1/2$ at the symmetric point $J_d=0$ but will not investigate them further.  
We also report the appearance, for the other plateaus, of several staggered spin imbalance phases whose order parameters are also quantized. 

%%%%%%%%%%%%%%%%%%%%%%%%%%%%%%%%%%%%%%%%%%%%%%%%%%%%%%%%%%%%%%%%%%%%%%%%%%%%%%%%%%%%%%%%%%

\subsubsection{Highest plateau: \texorpdfstring{$J_d>0$}{Jd > 0}}

For $J_d>0$, as increasing $J_\parallel/J_\perp$, we expected the two-fold degenerate ground state with uniform tetramer order from Sec.~\ref{sec:SCHam_genS_Jdgreater}. 
For $J_d/J_\perp=1$, the transition point was estimated as $J_\parallel/J_\perp \simeq 0.35$ from Fig.~\ref{fig:ParametersXYZS1Jd05}. 
However we could not observe any sign of symmetry broken phase after the staggered spin imbalance order vanishes. 
A useful quantity to identify the critical behavior of the system is the von Neumann entanglement entropy of a block $S_{vN}(\ell)$, which exhibits two different behaviors for large block sizes $\ell$ under the OBC: 
$S_{vN}(\ell)$ saturates to a constant when the system is gapped, whereas $S_{vN}(\ell) \simeq (c/6) \log \ell +c'$ when the system is critical.~\cite{Calabrese2004} 
Here $c$ is the central charge of the underlying conformal field theory and $c'$ is a nonuniversal constant. 
Finite-size effects are correctly treated through the conformal map, $\ell \rightarrow d(\ell|L)=(L/\pi) \sin (\ell\pi/L)$. 

\begin{figure}
\begin{center}
\includegraphics[width=\columnwidth,clip]{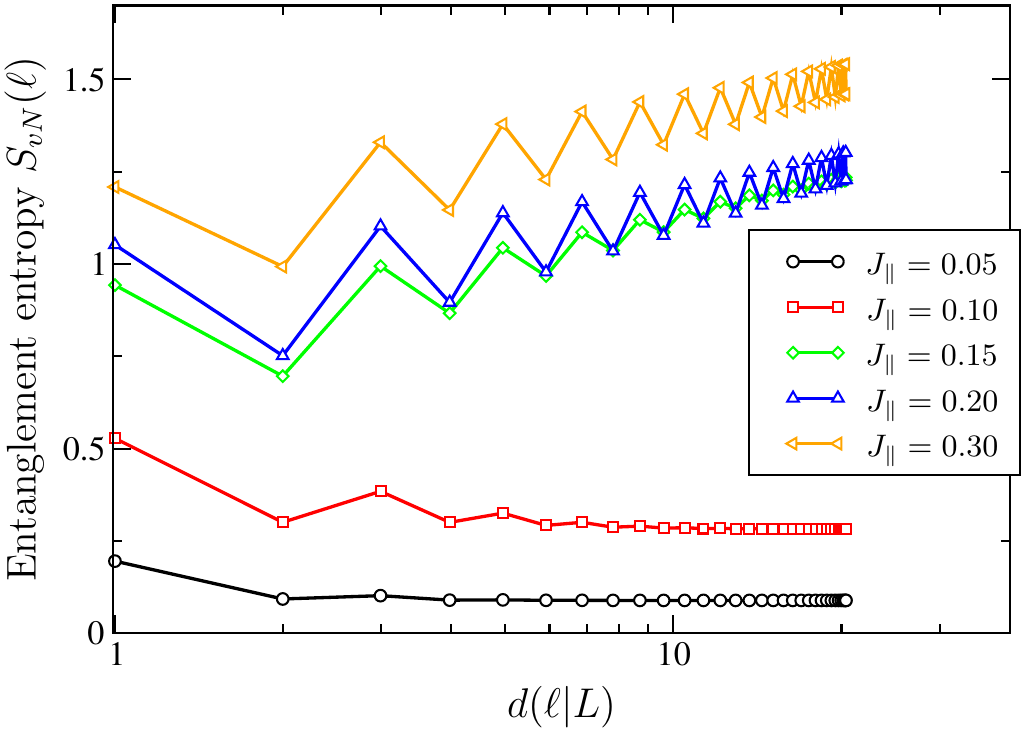}
\end{center}
\caption{(Color online) Evolution of the block entanglement entropy $S_{vN}(\ell)$ vs block length 
$d(\ell|L)$ (starting at one end of the tube) on a $L=64$ tube at $J_d/J_{\perp}=1$ when $J_\parallel/J_\perp$ is varied from 0.05 to 0.3.
The logarithmic scale is used for the horizontal axis.}
\label{fig:entropyS1_Jd1_QPT}
\end{figure}

In Fig.~\ref{fig:entropyS1_Jd1_QPT}, we plot the entanglement entropy for $J_d/J_{\perp}=1$ and various values of $J_\parallel/J_\perp$. 
Starting at $J_\parallel/J_\perp=0.05$, we observe the flat behavior of $S_{vN}$ in the spin imbalance phase with a finite gap. 
Around $J_{\parallel}/J_{\perp} = 0.15$, its behavior changes and a logarithmic fitting, after removing the oscillating part coming from a bond modulation~\cite{Capponi2013}, gives a central charge close to 1 for a wide range of $J_\parallel/J_\perp$ ($c=0.96,0.99,0.93$ for respectively $J_{\parallel}/J_{\perp}=0.15,0.2,0.3$). 
This does not agree with our expectation that another gapped phase with uniform tetramer order appears from the strong-coupling analysis. 
The wide critical phase with $c=1$ observed here is with no doubt a numerical artifact and was anticipated from the following reason: 
Since the difference between $J_y$ and $J_z$ in the effective XYZ model \eqref{eq:Ham_xyzS1_2} is very small ($\sim 0.04 J_\parallel^2/J_\perp$) at the second order, the tetramer phase dominated by $J_z$ is close to an easy-plane antiferromagnetic phase with $c=1$. 
Thus, the excitation gap should be very small and this means that in a numerical simulation we will find a critical behavior on system sizes smaller than the correlation length. 
We plotting the estimated central charge as a function of the coupling, we can still use the abrupt jump to locate the phase transition.

%%%%%%%%%%%%%%%%%%%%%%%%%%%%%%%%%%%%%%%%%%%%%%%%%%%%%%%%%%%%%%%%%%%%%%%%%%%%%%%%%%%%%%%%%%

\subsubsection{Highest plateau: \texorpdfstring{$J_d=0$}{Jd = 0}}

\begin{figure}
\begin{center}
\includegraphics[width=\columnwidth,clip]{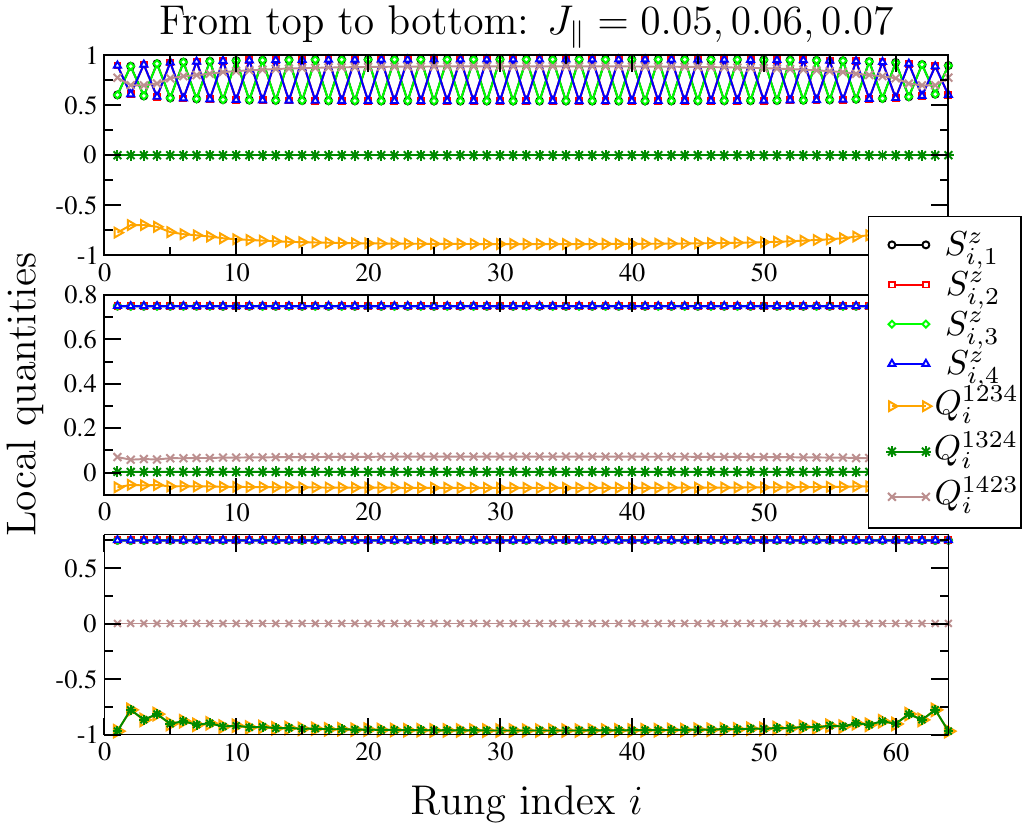}
\end{center}
\caption{(Color online) Local quantities $S^z_{i,j}$ and $Q^{jklm}_i$ as functions of the rung index $i$ at the symmetric point $J_d=0$, $L=64$ and $S=1$. 
The results are shown for $J_\parallel/J_\perp=0.05$, $0.06$, and $0.07$ from top to bottom.}
\label{fig:Jdeq0_loc_S1_qpt}
\end{figure}

At the $S_4$ symmetric point, we begin by showing in Fig.~\ref{fig:Jdeq0_loc_S1_qpt} the evolution of the local quantities $\langle S^z_i \rangle$ and $\langle Q_i^{jklm}\rangle$ with varying $J_\parallel/J_\perp$. 
We see that at $J_{\parallel}/J_{\perp}=0.05$ the staggered spin imbalance is present but starts to vanish, and is completely absent for $J_{\parallel}/J_{\perp}=0.06$ and larger values. 
This gives us a rough estimate of the transition and is in agreement with the value expected from the ED calculation on the effective model (see Fig.~\ref{fig:ED_Heff_GenS} (a) where the excited levels start to collapse on the three-fold degenerate ground state at $J_\parallel/J_\perp \simeq 0.06$). 
Also, we can compare the expectation values of the $Q$ operators to the values given in the Table~\ref{tab:Q_averages} for $S=1$. 
We see that the simulation for $J_{\parallel}/J_{\perp}=0.07$ selects the $| \Psi_{-} \rangle$ state, and that the expectation value of the $Q$ operators are almost the halves of those in the ideal tetramer states. 
From the Hamiltonian \eqref{eq:Heff_2nd}, even if the $q_3$ term dominates and causes the tetramer order, the other terms with $q_1$ and $q_2$ are still not negligible in the sense that they give the quantum fluctuations around this state. 
This is different from the spin imbalance order, where the order parameter gives the almost quantized value, indicating the strong suppression of quantum fluctuation. 
Those ideas will be developed more deeply in the conclusion.

\begin{figure}
\begin{center}
\includegraphics[width=\columnwidth,clip]{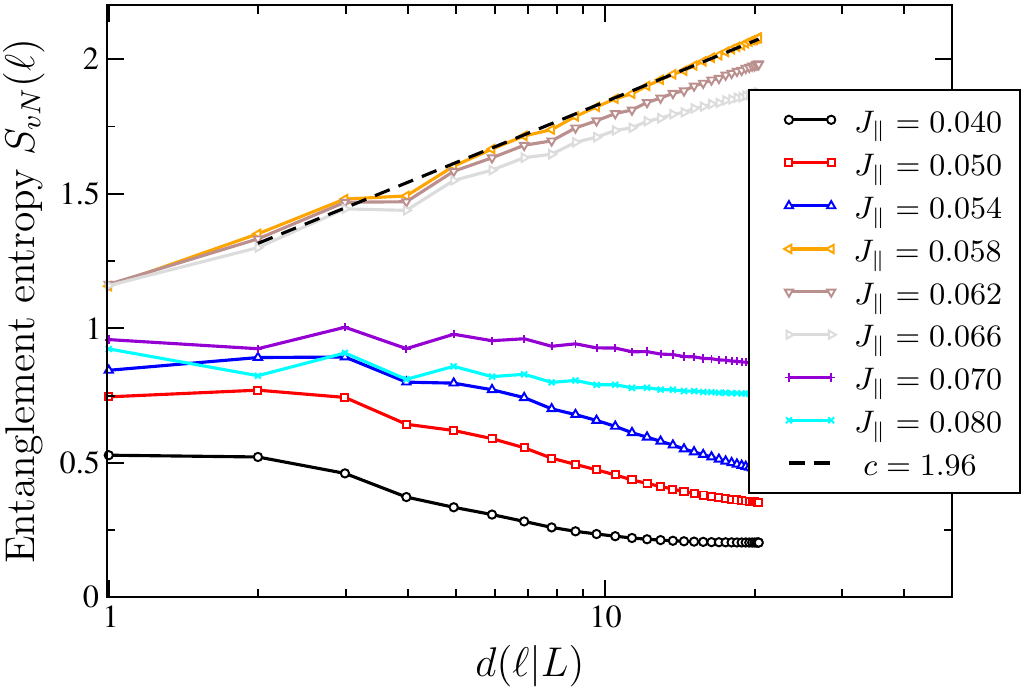}
\end{center}
\caption{(Color online) Evolution of the block entanglement entropy $S_{vN}(\ell)$ vs block length $d(\ell|L)$ (starting at one end of the tube) on a $L=64$ tube at $J_d/J_{\perp}=0$ when $J_{\parallel}$ is varied from $0.04$ to $0.08$. 
The logarithmic scale is used for the horizontal axis. 
The entropy for $J_{\parallel}/J_{\perp}=0.058$ is well fitted by the logarithmic function $(c/6) \log d(\ell|L) +c'$ with $c=1.96$, indicated by the dashed line. }
\label{fig:entropyS1_Jd0_QPT}
\end{figure}

Then, we use the entanglement entropy to precisely locate the phase transition. 
In Fig.~\ref{fig:entropyS1_Jd0_QPT}, we plot the entanglement entropy for several values of $J_\parallel/J_\perp$ around the transition point. 
The saturated behavior on both sides of the transition confirms the gapped phases, and we see that for $J_{\parallel}/J_{\perp}=0.058$ the von Neumann entropy is logarithmically fitted with a central charge $c=1.96$, indicating some exotic criticality. 
The question is then whether this value is trustworthy or not. 
This $c=2$ could point towards the criticality governed by the level-1 SU(3) Wess-Zumino-Witten model. 
Neglecting the next-nearest-neighbor terms in the effective model \eqref{eq:Heff_2nd}, the model could be at or in the vicinity of such criticality (an exact SU(3) symmetric point is at $q_1=q_2=q_3$). 
However, we could not find any evidence of the criticality with $c=2$. 
As in the case for $J_d>0$, even though a microscopic Hamiltonian does not possess the exact symmetry, the effective continuum theory at the transition may exhibit the emergent symmetry. 
We believe that this result could also be a numerical artifact, maybe signaling the presence of some critical point in the vicinity of our model. 
For larger system sizes this critical behavior could be replaced by a first-order transition as was argued for instance in Ref.~\onlinecite{Charrier2010}.

%%%%%%%%%%%%%%%%%%%%%%%%%%%%%%%%%%%%%%%%%%%%%%%%%%%%%%%%%%%%%%%%%%%%%%%%%%%%%%%%%%%%%%%%%%

\subsubsection{Other plateaus: quantized spin imbalance phases}

\begin{figure}
\begin{center}
\includegraphics[width=\columnwidth,clip]{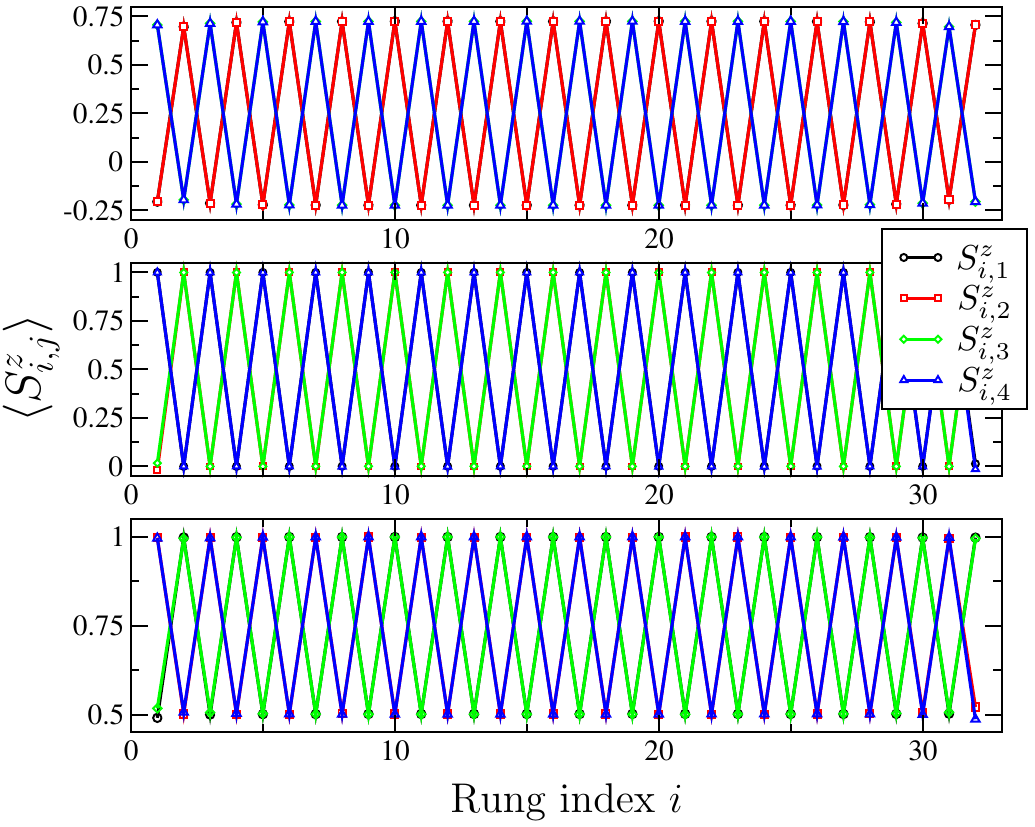}
\end{center}
\caption{(Color online) Local magnetization $\langle S^z_{i,j} \rangle$ at the symmetric point $J_d=0$, $J_{\parallel}/J_{\perp} = 0.01$, and $S=1$, on a $L=32$ tube. 
From top to bottom, the panels correspond to magnetization plateau $m=1/4,1/2,3/4$.}
\label{fig:SpinImbalance_S1}
\end{figure}

\begin{figure}
\begin{center}
\includegraphics[width=\columnwidth,clip]{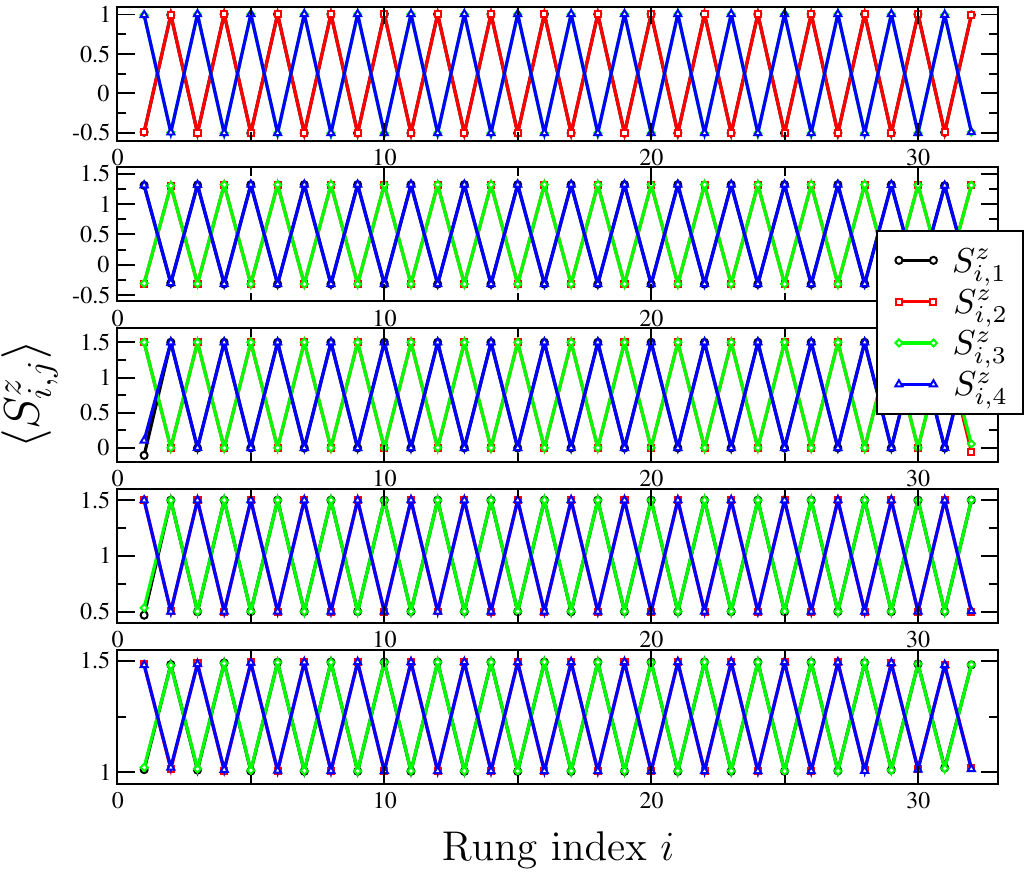}
\end{center}
\caption{(Color online) Local magnetization $\langle S^z_{i,j} \rangle$ at the symmetric point $J_d=0$, $J_{\parallel}/J_{\perp} = 0.01$, and $S=3/2$, on a $L=32$ tube. 
From top to bottom, the panels correspond to magnetization plateau $m=1/4,1/2,3/4,1,5/4$.}
\label{fig:SpinImbalance_S32}
\end{figure}

We end this section by plotting in Fig.~\ref{fig:SpinImbalance_S1} the local magnetization $\langle S^z_{i,j} \rangle$ computed at the symmetric point $J_d=0$, $J_{\parallel}/J_{\perp} = 0.01$, and $S=1$, on the magnetization plateaus $m=1/4,1/2,3/4$ from top to bottom for $L=32$. 
Fig.~\ref{fig:SpinImbalance_S32} shows the local magnetization for $S=3/2$ on the magnetization plateaus $m=1/4,1/2,3/4,1,5/4$ from top to bottom for the same parameters.

All the plateaus display the presence of a staggered spin imbalance with a quantized value $\langle \mu_i^{jklm} \rangle \in \mathbb{Z}$ (with the appropriate choice of the indices $(jklm)$ depending on which ground state is selected), aside from a small discrepancy for the $m=1/2$ plateau for $S=3/2$ in Fig.~\ref{fig:SpinImbalance_S32}. 
As said before, our strong-coupling analysis is only available for the highest plateaus (in the present case, $m=3/4$ for $S=1$ and $m=5/4$ for $S=3/2$). 
But one can remark that starting from the highest plateau where two spins are polarized to $+S$ and the two others have a magnetization $S-1/2$, the pattern on the next lower plateau is given by simply decreasing this last value by $1/2$. 
This holds for all the plateaus except the lowest one, for both $S=1$ and $S=3/2$. 
The former basically suggests that those states can be understood at the mean-field level by simply minimizing the diagonal term $S^z_{i,j}S^z_{i+1,j}$ of the longitudinal coupling, thus the ground states would be once again direct-product states over the rungs. 
For the two lowest plateaus not following this pattern, we did not find any simple explanation for the computed local magnetizations. 
Finally, the figures also indicate that the degeneracy of the ground state for each plateau should be identical to what we obtained previously, namely six at the symmetric point.

We do not show the magnetization profiles for $J_d>0$ because it is in fact trivial. 
A quick reasoning on coupling the four spins in a single tetrahedron tells us that if the plateau has an even total $S^z_{\boxtimes}$, then the ground state is unique and no ordered phase will be present, and if it is odd the ground state is two-fold degenerate. 
When coupling the tetrahedra, the perturbation theory always leads to the Ising model (\ref{eq:Ham_tauxtaux}) and as a consequence, the spin imbalance amplitude is always 
 minimal as for $S=1/2$ and $J_d>0$. 
For example, for $S=1$, the Ising effective Hamiltonian for the plateau with $m=1/4$ ({\it i.e.} $S^z_{\boxtimes}=1$) displays a staggered spin imbalance characterized by $\langle S^z_{1,3} \rangle =1/2$ and $\langle S^z_{2,4} \rangle =0$ and conversely on the neighboring rungs (data not shown). This also means that for those even plateaus, there is a discontinuity between $J_d>0$ and the symétric point, for which we have seen spin imbalance phases on all the plateaus.

%%%%%%%%%%%%%%%%%%%%%%%%%%%%%%%%%%%%%%%%%%%%%%%%%%%%%%%%%%%%%%%%%%%%%%%%%%%%%%%%%%%%%%%%%%
%%%%%%%%%%%%%%%%%%%%%%%%%%%%%%%%%%%%%%%%%%%%%%%%%%%%%%%%%%%%%%%%%%%%%%%%%%%%%%%%%%%%%%%%%%
%%%%%%%%%%%%%%%%%%%%%%%%%%%%%%%%%%%%%%%%%%%%%%%%%%%%%%%%%%%%%%%%%%%%%%%%%%%%%%%%%%%%%%%%%%

\subsection{Relation to the path-integral results} \label{sec:comp_path_int}

In the $S=1/2$ strong-coupling approach of Sec.~\ref{sec:Strong_coupling}, we predicted the stabilization of 
staggered spin imbalance phases in the regime of small $J_{\parallel}/J_{\perp}$. Its quantized magnitude was
understood in terms of the hidden ferromagnetism through the nonlocal transformation. For higher values of $S$, the perturbation theory becomes too involved 
because of the large number of low-energy states to take into account. But, the tetramer ground state for the highest plateau has the same form as for $S=1/2$, 
and we could repeat our analysis. It led to the prediction of a second phase on the plateau, characterized by a ferromagnetic ordering of the $Q$ operator (\ref{eq:rung_ops}). All those predictions
were confirmed numerically by using DMRG simulations. We also reported the observation of quantized staggered spin imbalance
phases for the other plateaus, although we lack of an effective theory to understand them.

We want to make the connection with the
path-integral results. Our semi-classical approach is able to predict $k_{\parallel}=0$ orders and can indeed describe the tetramer ordered phase. From the discussion
of Sec.~\ref{sec:PI_discussion}, this phase corresponds to the free energy with minima at $\alpha^*=0,\pi$ computed for moderately small values of $J_{\parallel}$, with the absence of tunneling
between the two wells, i.e. the $\mathbb{Z}_2$ symmetry is broken. The degeneracies for 
both $J_d=0$ (three) and $J_d>0$ (two) as well as the predicted $k_{\parallel}=0$ 
ordering of $Q$ match with our numerical results. On the other hand, 
as we previously explained the staggered phase cannot be recovered in our calculation. Yet, the same mechanism proposed for a uniform quantized spin imbalance in terms of
the delocalization of an angular field is at play here.

We mentioned that in order to describe $k_{\parallel}=\pi$ phases in the semi-classical approach, it is necessary to double the unit cell. 
This is done by considering the spin operators (\ref{eq:PI_spin_operators}) on two sublattices $A$ and $B$ on every 
chain and consequently working with eight fields $\Pi_{i,p}$ where $i=1,...,4$ labels the chain and $p=A,B$ labels 
the sublattice, and similarly the angular variables $\varphi_{i,p}$. Thus in the calculation we would have to consider, 
after the transformation (\ref{eq:PI_Pi_transformation}), the $\Omega_{i,p}$ on the two sublattices. We can then 
construct the homogeneous and staggered fields $\Omega_{i,h/s}=\Omega_{i,A} \pm \Omega_{i,B}$. The numerical data
clearly indicates that $\Omega_{2,s}$ is locked to its (nonzero) eigenvalues, and therefore 
its angular conjugate $\phi_{2,s}$ is necessarily delocalized. This is also confirmed by the effective Hamiltonian. From the Ising model at $J_d>0$,
the spin imbalance ground state is expected to be very close to a product state $| \Psi \rangle= \bigotimes_{i \in A, i' \in B} |\Psi_{13} \rangle_i |\Psi_{24} \rangle_{i'}$ (or the one obtained by interchanging $A$ and $B$). We computed 
the entanglement entropy between two rungs in the DMRG simulations and indeed found a value very close to zero (data not shown). Combined with 
the plaquette states given in Eq.~(\ref{eq:Psi_singlets}), it ensures that the fluctuation of the spin imbalance (of $\Omega_{2,s}$) are suppressed, confirming that 
the field is strongly pinned to one value. The same argument tells us that $\Omega_{2,h}$, which takes a zero expectation value, has no fluctuation, thus $\Omega_{2,h}$ is also locked but to
its zero eigenvalue. This situation is different from the tetramer ordered phase coming from the XYZ model, where again $\langle \Omega_{2,h} \rangle = 0$ but 
can fluctuate. It is coherent with the interpretation of this phase that we gave above in terms of the broken $\mathbb{Z}_2$ ($\mathbb{Z}_3$) symmetry for $J_d>0$ ($J_d=0$), 
associated to the absence of tunneling between the minima and therefore of winding. In the
three-leg spin-tube, the staggered spin imbalance comes from an operator $\tau^x_{i-2}\tau^x_{i-1}\tau^x_{i}\tau^x_{i+1}\tau^x_{i+2}\tau^x_{i+3}$ perturbing an XXZ model in the bosonization picture~\cite{Okunishi2012}, and 
in that case nothing prevents the associated staggered $\Omega_{i,s}$ field to fluctuate.

We also briefly numerically studied the various magnetization plateaus for $S=1$ and $S=3/2$ at very small longitudinal coupling. 
It revealed the presence, for every plateau, of $k_{\parallel}=\pi$
spin imbalance phases (see Fig.~\ref{fig:SpinImbalance_S1} and Fig.~\ref{fig:SpinImbalance_S32}) displaying the same quantization phenomenon. Even though we cannot rely on
an effective Hamiltonian, we noticed that, except for two plateaus, two of the spins are always polarized to $+S$ while the magnetization of the two others
progressively decreases as $m$ is lowered. It indicates that the ground state can also be written as a product state and that the fluctuation of the spin imbalance is again
strongly suppressed.

It is however not evident how this delocalization happens if we start from the model (\ref{eq:4leg_tube_Ham}). Physically, it is obvious that the 
$\phi_{2,s}$ field should pick up a mass term $m^2_{2,s}\propto J_{\parallel}$ and a straightforward calculation with the doubled unit cell confirms it. A possible explanation is that
the staggered state observed is in fact representative of a modified Hamiltonian, and the model we study here is not sufficient in the semi-classical approach. 
On the other hand, one can see that adding an extra term 
$H_{\parallel}^{'}=J_{\parallel}^{'}\sum_{i=1}^L\sum_{j=1}^4 \vec{S}_{i,j} \cdot \left( \vec{S}_{i+1,j-1} + \vec{S}_{i+1,j+1} \right)$ in the Hamiltonian, {\it i.e.} including longitudinal
``twisted'' couplings, would favour the homogeneous spin imbalance. A quick calculation for $J_d>0$ shows that the factor in front of the Ising
model (\ref{eq:Ham_tauxtaux}) is changed to $J_{\parallel}-J^{'}_{\parallel}$. This new term adds more frustration to the problem and the two spin imbalance phases compete. In the
semi-classical approach, its effect is in particular to reduce the mass of the $\Omega_2$ term in the action (\ref{eq:PI_action2}) as $\lambda_2=4a(J_{\parallel}-J_{\parallel}^{'})+2aJ_d$. As this
prefactor is reduced, it is likely that the Gaussian order is not sufficient to capture correctly the behavior of the $\Omega_2$ field. Beyond our second-order 
calculation, new terms are expected to appear and favor the pinning of the field to nonzero values.

As a final remark, it is worth highlighting the direct link between the continuous degeneracy present at the classical level and its consequences on the quantum system.
For a generic unfrustrated system, in the path-integral formulation the presence of a magnetization plateau is explained by the delocalization of the angular field 
representing the Goldstone mode (the $\phi_4$ field here). As a consequence, its conjugate is locked (to zero) and a plateau appears. Usually, 
this scenario is not expected to happen for other fields. In the model studied here, we have the unusual situation in which soft modes, 
typical of highly frustrated systems, behave in some sense as supplementary Goldstone modes, although strictly speaking they are not.
Despite the fact that no symmetry protects them to have a localizing potential, frustration can make this effective potential weak enough to be overcome 
by tunneling effects. Then, these soft or pseudo Goldstone modes can experience a proliferation of vortices, as for the real Goldtsone mode, 
and delocalize. This has the effect of pinning the conjugate variable, which in the case at hand is directly related to the observed spin imbalance as we explained above.  
The relation between a magnetization plateau and the delocalization of the Goldstone mode is in fact not at all specific 
to one dimensional systems.~\cite{Tanaka2009} Neither is the presence of soft modes (within the spin wave description) 
in frustrated magnets, as it is for example the case for the Kagom\'{e} lattice~\cite{Chalker1992}. We have then good reasons 
to expect a similar kind of behavior in higher dimensional frustrated magnetic models, with the same phenomenology involved, that 
is, locking of spin imbalance and a ground state wave function which is very close to a product state. In that sense, the model 
studied here is a very good representative example of a wide family of highly frustrated magnets which present a very particular 
manifestation of the classical order by disorder at the quantum level, which goes beyond the most intuitive expectation, 
namely, a selection mechanism similar to the classical case. 

\section{Conclusion}\label{sec:conclusion}

In this paper, we have studied the behavior of a frustrated four-leg spin tube under a magnetic field. 
As expected, the system shows the presence of magnetization plateaus for a wide range of parameters. 
We have focused on the behavior of the system at the magnetization plateaus as it presents an interesting behavior that can be traced back to the presence of frustration. 
We used a combination of a path-integral approach, the analysis of strong-coupling effective Hamiltonians, and the DMRG method. 
The numerical results from DMRG show two intriguing properties of the ground state when sitting on the magnetization plateaus, namely: 
(i) The appearance of a spin imbalance which is locked to integer values, (ii) an almost perfect factorization of the ground state wave function. Moreover, we expect that the property (i) is a consequence of (ii). 

In the highest plateau, where the number of nonmagnetic degrees of freedom per rung is small enough, a relatively simple low-energy effective Hamiltonian can be constructed. 
The analysis of the effective Hamiltonian confirms the behavior described above. 
It is interesting to notice that at the most frustrated point, the effective Hamiltonian calculated to first order is equivalent, via a nonlocal transformation, to a spin-$1$ ferromagnetic Heisenberg chain. 
The macroscopic degeneracy of the ground state is lifted by the higher-order corrections to the effective Hamiltonian. 
It is, in principle, not evident at all that the higher-order corrections give rise to the (almost) factorization properties of the corresponding six-fold degenerate ground state. 
This can be seen from the fact that the staggered correlation function, corresponding to a string correlation function via the nonlocal transformation, is almost saturated, indicating indeed a factorized structure. 

Last but not least, we would like to insist that, in fact, the above scenario is also reproduced for other plateaus where the effective Hamiltonian is more complicated, because of the presence of more low-energy (non-magnetic) degrees of freedom. 
Although no tractable effective Hamiltonian is available in the general case, it can be seen in the numerical results and argued as a delocalization of a pseudo Goldstone mode corresponding to the canonical conjugate variable to the spin imbalance. 
States having the properties (i) and (ii) were already shown to be exact ground states in a wide variety of frustrated systems \cite{Schulenburg2002}, but in fact this property remains almost intact to a very large extent even when the ground state cannot be obtained exactly.~\cite{Capponi2013} 
Increasing further the magnetization on those systems may either imply a jump in the magnetization curve (as it happen in Ref.~\onlinecite{Schulenburg2002}) or simply a delocking of the spin imbalance which is due to the onset of quasi-long-range order (or just simply long-range order in higher dimensions) associated with the true Goldstone mode enforcing itself a localization of the pseudo Goldstone mode conjugate to the spin imbalance.

\begin{acknowledgments}
We thank M. Oshikawa for fruitful discussions. 
Y.F. was supported in part by the Program for Leading Graduate Schools, MEXT, Japan. 
Numerical simulations were performed at CALMIP and GENCI.
Y.F. also thanks the hospitality of Laboratoire de Physique Th\'eorique, UPS and CNRS, where this work was completed. 
\end{acknowledgments}

%%%%%%%%%%%%%%%%%%%%%%%%%%%%%%%%%%%%%%%%%%%%%%%%%%%%%%%%%%%%%%%%%%%%%%%%%%%%%%%%%%%%%%%%%%%%%%%%%%%%%%%%%%%%
%%%%%%%%%%%%%%%%%%%%%%%%%%%%%%%%%%%%%%%%%%%%%%%    APPENDIX    %%%%%%%%%%%%%%%%%%%%%%%%%%%%%%%%%%%%%%%%%%%%%%
%%%%%%%%%%%%%%%%%%%%%%%%%%%%%%%%%%%%%%%%%%%%%%%%%%%%%%%%%%%%%%%%%%%%%%%%%%%%%%%%%%%%%%%%%%%%%%%%%%%%%%%%%%%%
\appendix

\section{Notes on strong-coupling expansion} \label{sec:appendixA}

\subsection{Gell-Mann matrices} \label{sec:GellMann}

A convention of the Gell-Mann matrices used in this paper are given by 
\begin{eqnarray}
& \lambda^1 = \left( \begin{array}{ccc} 0 & 1 & 0 \\ 1 & 0 & 0 \\ 0 & 0 & 0 \end{array} \right), \hspace{10pt} 
\lambda^2 = \left( \begin{array}{ccc} 0 & -i & 0 \\ i & 0 & 0 \\ 0 & 0 & 0 \end{array} \right), & \nonumber \\
& \lambda^3 = \left( \begin{array}{ccc} 1 & 0 & 0 \\ 0 & -1 & 0 \\ 0 & 0 & 0 \end{array} \right), \hspace{10pt} 
\lambda^4 = \left( \begin{array}{ccc} 0 & 0 & 1 \\ 0 & 0 & 0 \\ 1 & 0 & 0 \end{array} \right), & \nonumber \\ 
& \lambda^5 = \left( \begin{array}{ccc} 0 & 0 & -i \\ 0 & 0 & 0 \\ i & 0 & 0 \end{array} \right), \hspace{10pt} 
\lambda^6 = \left( \begin{array}{ccc} 0 & 0 & 0 \\ 0 & 0 & 1 \\ 0 & 1 & 0 \end{array} \right), & \nonumber \\ 
& \lambda^7 = \left( \begin{array}{ccc} 0 & 0 & 0 \\ 0 & 0 & -i \\ 0 & i & 0 \end{array} \right), \hspace{10pt} 
\lambda^8 = \frac{1}{\sqrt{3}} \left( \begin{array}{ccc} 1 & 0 & 0 \\ 0 & 1 & 0 \\ 0 & 0 & -2 \end{array} \right). &
\end{eqnarray}

\subsection{Nonlocal transformation of the strong-coupling Hamiltonian} \label{sec:hidden_sym}

We here explain the hidden SU(2) symmetry in the first-order effective Hamiltonian in Sec.~\ref{sec:SCHam} under the OBC. 
As shown by Kennedy~\cite{Kennedy1994}, any spin-1 Hamiltonian with short-range interactions, $T^\alpha_i T^\alpha_{i+1}$ and $(T^\alpha_i)^2$, 
can be mapped onto some Hamiltonian written in terms of short-range bilinear interactions of three anticommuting operators by a nonlocal unitary transformation. 
One can easily see that $\lambda^{1,4,6}$ satisfy the anticommutation relation $\left\{ \lambda^\mu_i, \lambda^\nu_i \right\} =\lambda^\rho_i$ where $(\mu,\nu,\rho)$ is any permutation of $(1,4,6)$. 
Since the first-order effective Hamiltonian~\eqref{eq:Heff_1st} is precisely in the latter form, we can conversely use his result and obtain some spin-1 Hamiltonian. 
The desired nonlocal unitary transformation has been proposed in Ref.~\onlinecite{Kennedy1994} and written as a product of two unitary operators 
$\mathcal{V} =\mathcal{U}_\textrm{KT} \mathcal{W} $ where $\mathcal{U}_\textrm{KT}$ is the nonlocal one found by Kennedy and Tasaki~\cite{Kennedy1992,Oshikawa1992}, 
\begin{eqnarray}
\mathcal{U}_\textrm{KT} = \prod_{j<k} \exp \left( i\pi T^z_j T^x_k \right), 
\end{eqnarray}
and $\mathcal{W}$ is a product of local unitary operators,
\begin{eqnarray}
\mathcal{W} = \prod_k \mathcal{W}_k, \hspace{10pt} 
\mathcal{W}_k = \left( \begin{array}{ccc} 1/\sqrt{2} & 0 & 1/\sqrt{2} \\ 0 & 1 & 0 \\ 1/\sqrt{2} & 0 & -1/\sqrt{2} \end{array} \right). 
\end{eqnarray}
Under this transformation $\mathcal{V}$, the bilinear operators are transformed as 
\begin{equation} \label{eq:KT_transf_effect}
\begin{split}
& \mathcal{V} \lambda_i^1 \lambda_{i+1}^1 \mathcal{V}^{-1} = -T^x_i T^x_{i+1}, \\
& \mathcal{V} \lambda_i^4 \lambda_{i+1}^4 \mathcal{V}^{-1} = -T^z_i T^z_{i+1}, \\
& \mathcal{V} \lambda_i^6 \lambda_{i+1}^6 \mathcal{V}^{-1} = -T^y_i T^y_{i+1}. 
\end{split}
\end{equation}
and the Hamiltonian~\eqref{eq:Heff_2nd} is exactly mapped onto the spin-1 Heisenberg ferromagnetic chain, 
\begin{eqnarray}
\mathcal{V} H^{(1)}_\textrm{eff} \mathcal{V}^{-1} = -\frac{J_\parallel}{4} \sum_{i=1}^L \vec{T}_i \cdot \vec{T}_{i+1}, 
\end{eqnarray}
under the OBC. 

Once we concern the second-order perturbation as in Eq.~\eqref{eq:Heff_2nd}, the transformed Hamiltonian no longer exhibits the exact SU(2) symmetry but still remains a local form. 
For instance, few of the additional terms are given by 
\begin{eqnarray}
&\calV \lambda^2_i \lambda^2_{i+1} \calV^{-1} = -(T^y_i T^z_i + T^z_i T^y_i) (T^y_{i+1} T^z_{i+1} + T^z_{i+1} T^y_{i+1}),& \nonumber \\
&\calV \lambda^3_i \lambda^3_{i+1} \calV^{-1} = \left[ (T^z_i)^2-(T^y_i)^2 \right] \left[ (T^z_{i+1})^2 - (T^y_{i+1})^2 \right],& \nonumber \\
&\calV \lambda^1_i \lambda^4_{i+1} \lambda^6_{i+2} \calV^{-1} = T^x_i (T^x_{i+1} T^y_{i+1} + T^y_{i+1} T^x_{i+1}) T^y_{i+2},& \nonumber \\
&\calV \lambda^1_i (1-\sqrt{3} \lambda^8_{i+1}) \lambda^1_{i+2} \calV^{-1} = 3 T^x_i \left[ (T^x_{i+1})^2-1 \right] T^x_{i+2}.& \nonumber \\
\end{eqnarray}
The effect of a finite diagonal coupling $J_d \neq 0$ in Eq.~\eqref{eq:Heff_asym} is written as 
\begin{eqnarray}
\calV \lambda^8_i \calV^{-1} = -\frac{1}{\sqrt{3}}+\sqrt{3}(T^x_i)^2. 
\end{eqnarray}

%\bibliography{BiblioSpinTube}
%\bibliographystyle{./apsrev4-1}

%%%%%%%%%%%%%%%%%%%%%%%%%%%%%%%%%%%%%%%%%%%%%%%%%%%%%%%%%%%%%%
%     BIBLIOGRAPHY 

\end{document}